\documentclass[aps,prb,twocolumn,amsmath,amssymb,superscriptaddress,reprint,floatfix]{revtex4-2}
\usepackage{upgreek}
\usepackage{graphicx}
\usepackage{commath}
\usepackage{bm}
\usepackage{dsfont}
\usepackage[usenames,dvipsnames]{xcolor}
\usepackage{pstricks}
\usepackage[tight]{subfigure}
\usepackage{verbatim}
\usepackage{units}
\usepackage{multirow}
\usepackage{mathrsfs}
\usepackage{leftidx}
\usepackage{xspace}
\usepackage{braket}
\usepackage{bbm}
\usepackage{amsfonts,amssymb,amsmath}
\usepackage{psfrag}
\usepackage[abs]{overpic}
\usepackage{hyperref}
\hypersetup{colorlinks=true,linktoc=all,linkcolor=blue,breaklinks=true,citecolor=blue,urlcolor=blue}

\usepackage{lineno}  

\usepackage{booktabs}
\usepackage{mathtools}
\usepackage[version=4]{mhchem}
\usepackage{makecell}

\usepackage[labelfont=bf,labelsep=period]{caption}
\usepackage{algorithm}
\usepackage{algpseudocode}
\renewcommand{\thealgorithm}{\arabic{algorithm}}

\makeatletter
\newenvironment{myalgorithm}[2][t]
{%
    \refstepcounter{algorithm}%
    \begin{figure}[#1]
    \small
    \hrule height 0.8pt
    \vspace{4pt}
    \centering
    \textbf{Algorithm~\thealgorithm}\ #2\par
    \vspace{4pt}
    \hrule height 0.5pt
    \vspace{4pt}
    \raggedright
    \begin{minipage}{\columnwidth}
    \begin{algorithmic}[1]
}
{%
    \end{algorithmic}
    \end{minipage}
    \vspace{4pt}
    \hrule height 0.8pt
    \end{figure}
}
\makeatother

\begin{document}

\title{KappaFormer: Physics-aware Transformer for lattice thermal conductivity via cross-domain transfer learning}

\author{Mengfan Wu}
\altaffiliation{These authors contributed equally to this work}
\affiliation{Shanghai Research Institute for Intelligent Autonomous Systems $\&$ School of Physics Science and Engineering, Tongji University, Shanghai 200092, China}
\affiliation{Center for Phononics and Thermal Energy Science, Tongji University, Shanghai 200092, China}

\author{Junfu Tan}
\altaffiliation{These authors contributed equally to this work}
\affiliation{School of Electrical and Computer Engineering, Georgia Institute of Technology, Atlanta 30332, USA}

\author{Yu Zhu}
\affiliation{Department of Chemistry, Fudan University, Shanghai 200438, China}

\author{Jie Ren}
\email [Corresponding author\\E-mail: ]{Xonics@tongji.edu.cn}
\affiliation{Shanghai Research Institute for Intelligent Autonomous Systems $\&$ School of Physics Science and Engineering, Tongji University, Shanghai 200092, China}
\affiliation{Center for Phononics and Thermal Energy Science, Tongji University, Shanghai 200092, China}

\date{\today}

\begin{abstract}
Machine learning has been widely used for predicting material properties. However, efficient prediction of lattice thermal conductivity ($\kappa_\mathrm{L}$) remains a long-standing challenge, primarily due to the scarcity of high-quality training data. Here we introduce KappaFormer, a physics-aware Transformer architecture that embeds the harmonic–anharmonic decomposition of $\kappa_\mathrm{L}$ within the network. KappaFormer comprises a harmonic branch pre-trained on large-scale elastic property data and an anharmonic branch fine-tuned on limited experimental $\kappa_\mathrm{L}$ data, enabling effective knowledge transfer and enhanced generalization. High-throughput screening with KappaFormer identifies multiple candidates with ultralow $\kappa_\mathrm{L}$, which are further confirmed by first-principles calculations. Physics interpretability further elucidates the vibrational mechanisms governing thermal transport suppression, linking structural motifs to strong anharmonicity. This study provides a generalizable framework for physics-guided machine learning to accelerate the discovery of new materials.
\end{abstract}

\maketitle
\section{Introduction}  
In recent years, machine learning (ML) has significantly transformed the research paradigm in materials science.\cite{butler2018machine, merchant2023scaling, lookman2026materials, ahlawat2026family, li2023exploiting} As a powerful technology of artificial intelligence (AI), it has substantially reduced computational costs while expanding the explored boundaries of chemical space, spanning applications from forward screening driven by target property predictions to inverse design enabled by deep generative models,\cite{cheng2026artificial, guo2025generative, li2025probing} which cover diverse material systems such as thermoelectric materials,\cite{wang2024interpretable, wu2023machine} porous materials,\cite{wu2025ai, hu2025machine} battery materials,\cite{lu2024crystal} etc. This progress is largely driven by the rapid advancement of ML algorithms and high-performance computing. In addition, the open-access material databases also provide essential foundations for ML studies in materials science, such as the Materials Project (MP),\cite{jain2013commentary} Open Quantum Materials Database (OQMD),\cite{saal2013materials, kirklin2015open} Automatic-FLOW for materials discovery (AFLOW),\cite{curtarolo2012aflow} and Joint Automated Repository for Various Integrated Simulations (JARVIS),\cite{choudhary2020joint} etc.

Particularly, property predictions targeting well-defined physical properties enable high-throughput screening and quantitative evaluation within a broad chemical space, thereby directly facilitating function-oriented materials design.\cite{xie2018crystal, dong2025accurate, li2024high, li2024high, wu2025hierarchy, cui2023atomic} By serving as efficient surrogate models for computationally expensive atomistic simulations, ML algorithms based on graph neural networks,\cite{wu2020comprehensive} equivariant neural networks,\cite{satorras2021n, liao2023equiformer, equiformer_v2} and Transformer\cite{vaswani2017attention, ying2021transformers} architectures have achieved high predictive accuracy for many key material properties, including electronic,\cite{li2022deep} thermal,\cite{riebesell2025framework} mechanical,\cite{wen2024equivariant} and optical performance.\cite{klimova2025symmetry}
However, many important physical properties are not simple mappings from crystal structures, but instead arise from the complex interplay of multiple physical mechanisms across different aspects. This is especially true for complex properties such as lattice thermal conductivity ($\kappa_\mathrm{L}$), which depends on the collective behavior of lattice harmonicity and anharmonicity.\cite{chen2025softening, song2026role, wang2024revisiting} Purely end-to-end data-driven models must implicitly learn such relationships within high-dimensional feature spaces. This not only weakens the physical interpretability of ML models but also reduces data efficiency, thereby limiting the generalization ability in scenarios where experimental data are limited.

Materials with ultralow/high $\kappa_\mathrm{L}$ are essential for a wide range of technological applications, including energy sustainable development,\cite{qian2021phonon} thermal barrier coatings,\cite{padture2002thermal} and thermal management,\cite{song2018two, xiang2022thermal} etc. Particularly, with respect to energy sustainability, thermoelectric (TE) materials possessing ultralow $\kappa_\mathrm{L}$ facilitate the effective harvesting of waste heat from solar energy, manufacturing processes, and electronic systems for electricity generation. The conversion efficiency is theoretically determined by the dimensionless figure of merit ($zT$), which is defined as $zT={S^{2}\sigma}T/({\kappa_\mathrm{e}} + \kappa_\mathrm{L})$, where $S$ is the Seebeck coefficient, $\sigma$ is the electrical conductivity, $T$ is the absolute temperature, and $\kappa_\mathrm{e}$ is the electronic thermal conductivity, respectively. Given the intrinsic trade-offs among electronic transport properties, pushing $\kappa_\mathrm{L}$ to the ultralow limit of the thermal insulators represents an effective strategy for achieving optimal $zT$ values.\cite{snyder2008complex, nolas1999skutterudites, snyder2004disordered}

The conventional search for materials with desirable $\kappa_\mathrm{L}$ primarily relies on experimental trial-and-error approaches, remaining strongly dependent on advanced instrumentation and sample quality.\cite{balandin2011thermal, wei2016intrinsic} Advances in high-performance computing have enabled high-fidelity calculations of $\kappa_\mathrm{L}$ within the Boltzmann transport equation (BTE) framework,\cite{ShengBTE_2014, li2025high} which nevertheless demands the evaluation of harmonic and anharmonic properties at substantial computational costs.
Specifically, harmonic properties are determined from the second-order interatomic force constants (IFCs) and can be empirically approximated by elastic properties, whereas anharmonic properties are quantified through higher-order IFCs and are often evaluated empirically using the Grüneisen parameter $\gamma$.
Although many ML models have demonstrated promising performance in predicting $\kappa_\mathrm{L}$,\cite{jaafreh2021lattice, luo2023predicting, chen2019machine} their success is often limited by the scarcity of high-quality $\kappa_\mathrm{L}$ data, which is expensive to obtain from experiments. This challenge calls for models that can efficiently transfer learned knowledge under data-scarce scenarios. However, most existing methods adopt end-to-end learning strategies without explicitly incorporating the underlying physical principles of $\kappa_\mathrm{L}$, as is commonly observed in the property prediction models discussed above. Therefore, an interpretable predictive framework that combines efficient transfer learning with the preservation of intrinsic physical insights governing $\kappa_\mathrm{L}$ is expected to be developed.

In this study, we propose a physics-aware Transformer architecture for accurately predicting $\kappa_{\mathrm{L}}$, namely KappaFormer, which incorporates the physical insights of harmonicity and anharmonicity into graph-based attention networks. Furthermore, an efficient cross-domain transfer learning strategy jointly using large-scale elastic property data and limited experimental $\kappa_{\mathrm{L}}$ data is introduced, significantly enhancing model generalization and predictive accuracy under data-scarce conditions. Leveraging the pretrained KappaFormer, we perform large-scale $\kappa_{\mathrm{L}}$ predictions on tens of thousands of materials within the database, identifying a series of semiconductors with ultralow $\kappa_{\mathrm{L}}$. Among them, three candidates of CsNb$_2$Br$_9$, Cs$_2$AgI$_3$, and Cs$_6$CdSe$_4$ are further validated by solving the phonon BTE, with model interpretability and DFT analysis revealing the underlying mechanisms of phonon thermal transport.

\section{Results}  
\subsection{Architecture of KappaFormer}
\begin{figure*}[htbp]
\centering
\includegraphics[width=1\textwidth]{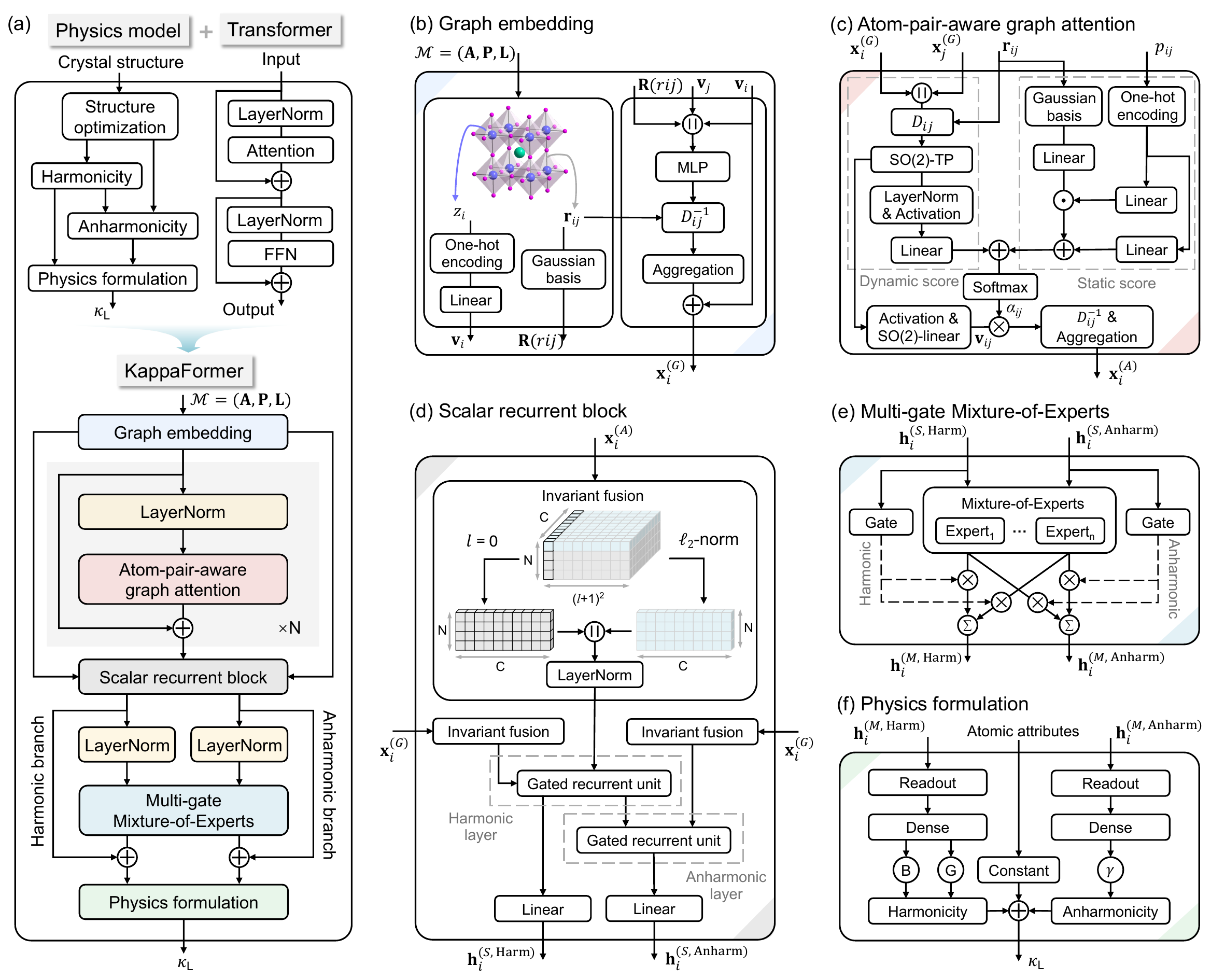}
\caption{\textbf{Architecture of KappaFormer.} a) Overview of the proposed framework that embeds the
harmonic–anharmonic decomposition of $\kappa_\mathrm{L}$ within a graph-based attention network. b) Graph embedding. c) Atom-pair-aware graph attention. d) Scalar recurrent block. e) Multi-gate Mixture-of-Experts. f) Physics formulation.}
\label{Fig1}
\end{figure*}

Inspired by the typical physics model for obtaining $\kappa_\mathrm{L}$ and the Transformer framework, we develop the architecture of KappaFormer as illustrated in Fig.~\ref{Fig1}a. On the one hand, $\kappa_{\mathrm{L}}$ can be calculated within a DFT-based workflow that begins with structure optimization, followed by the evaluation of harmonicity/anharmonicity-related properties, and then integration within a unified physical formulation. On the other hand, the Transformer framework employs attention mechanisms to capture interaction-aware feature representations from the input, followed by feed-forward networks (FFNs) that introduce nonlinear transformations to enhance output predictions. This correspondence between physics-driven decomposition and representation learning provides the foundation for KappaFormer, which embeds physically interpretable modules into a Transformer-based framework.

The KappaFormer model first encodes each periodic crystal as a multi-edge graph and applies atom-pair-aware graph attention as the backbone to learn the three-dimensional (3D) structural information, which greatly handles the geometric symmetries inherent in crystal structures. We then introduce the scalar recurrent block that operates on the equivariant embeddings of each structure, capturing the harmonicity and anharmonicity of the materials separately. These two branches are jointly interacted with each other through the Multi-gate Mixture-of-Experts (MMoE)\cite{ma2018modeling} to learn the physical dependencies between the harmonic and anharmonic properties. Finally, the physics formulation is integrated into downstream tasks to enable the accurate prediction of $\kappa_\mathrm{L}$ by explicitly modeling the harmonicity–anharmonicity decomposition, leading to improved out-of-distribution generalization. The details are described in the following subsections.


\subsubsection{Graph embedding}
Formally, a crystal structure can be described as $\mathcal{M} = (\mathbf{A}, \mathbf{P}, \mathbf{L})$, where $\mathbf{A}$ is the atomic type matrix, $\mathbf{P}$ is the position matrix, and $\mathbf{L}$ is the lattice matrix. And the crystal structure can be uniquely mapped onto a crystal graph $\mathcal{G} = (\mathcal{V}, \mathcal{E})$ embedded in 3D Euclidean space, where each vertex features $\mathbf{v}_i \in \mathcal{V}$ encodes the elemental information of an atom $i$, and each edge feature $\mathbf{e}_{ij} \in \mathcal{E}$ characterizes the bonding information between the atom pair $ij$ within a cutoff radius $r_{c}$ (Fig.~\ref{Fig1}b).

Each atom in the unit cell is associated with a vertex feature $\mathbf{v}_i$, which is initialized through an embedding layer that maps the atomic number $Z_{i}$ to a learnable feature representation
\begin{equation} \label{vertex}
   \mathbf{v}_i = \mathbf{W}_{v} \mathrm{OneHot}(Z_i).
\end{equation}
The edge features $\mathbf{e}_{ij}$ are constructed using a multi-edge graph under periodic boundary conditions. Each edge encodes both radial and directional information, where $\mathbf{r}_{ij}$ denotes the relative position vector from atom $i$ to atom $j$, and $r_{ij} = \|\mathbf{r}_{ij}\|$ is the corresponding interatomic distance. The radial component is expanded using a set of $M$ Gaussian basis functions as\cite{schutt2017schnet}
\begin{equation} \label{edge_1}
    R_m(r_{ij}) = \exp \left(- \frac{(r_{ij} - \mu_m)^{2}}{2\sigma^{2}} \right),
\end{equation}
and we denote the radial embedding as $\mathbf{R}(r_{ij}) = [R_1, R_2, \ldots, R_M]$. The radial embedding is then transformed via a learnable function implemented as a multilayer perceptron (MLP), which takes both radial and atomic features as input. The edge features are defined as
\begin{equation}  \label{edge_2}
    \mathbf{e}_{ij} = D^{-1}_{ij}(\mathbf{r}_{ij}) \cdot \left( \mathrm{MLP}\!(\mathbf{R}(r_{ij}) \,\Vert\, \mathbf{v}_i \,\Vert\, \mathbf{v}_j) \right)_{m=0}^{(l)},
\end{equation}
where the operator $\Vert$ denotes the concatenation of feature vectors, $D_{ij}$ denotes the Wigner-D matrix related to the relative position vector $\mathbf{r}_{ij}$, ensuring equivariance under rotations.\cite{equiformer_v2} 
The initial graph embedding for each atom $i$ is then given by
\begin{equation} \label{initial}
    \mathbf{x}_i^{(G)} = \mathbf{v}_i + \frac{1}{\alpha} \sum_{j \in \mathcal{N}(i)} \mathbf{e}_{ij},
\end{equation}
where $\alpha$ is a normalization factor, and $\mathcal{N}(i)$ denotes the set of neighboring atoms within a cutoff radius $r_c$ under periodic boundary conditions.

\subsubsection{Atom-pair-aware graph attention}
Since harmonicity and anharmonicity are intrinsically associated with pairwise force interactions, modeling atom–atom correlations at the attention level provides a natural inductive bias for resolving these physical contributions. An atom-pair-aware graph attention mechanism is introduced as shown in Fig.~\ref{Fig1}c, which is capable of capturing information from hidden embeddings and physical constraints. The $SO(2)$-tensor product (TP)\cite{equiformer_v2, escn} is employed to obtain the outputs $\mathbf{f}_{ij}$
\begin{equation} \label{so2}
    \mathbf{f}_{ij} = \mathcal{T} \left( D_{ij}(\mathbf{r}_{ij}) \cdot (\mathbf{x}_i^{(G)}\,\Vert\, \mathbf{x}_j^{(G)}), \mathrm{MLP}\!(\mathbf{R}(r_{ij}) \,\Vert\, \mathbf{v}_i \,\Vert\, \mathbf{v}_j) \right),
\end{equation}
which are split into two parts of irreducible representations (irreps) $\mathbf{f}_{ij}^{(L)}$ and scalar features $f_{ij}^{(0)}$, where $L$ denotes the maximum degree of the irreps. The value representation $\mathbf{v}_{ij}$ is obtained by applying a separable $S^2$ activation $\psi_S$ followed by the $SO(2)$ linear layer on $\mathbf{f}_{ij}^{(L)}$
\begin{equation} \label{value}
\mathbf{v}_{ij} =
SO_2 \text{-}\mathrm{Linear}\left(
\psi_S (\mathbf{f}_{ij}^{(L)})
\right).
\end{equation}

Here, two score mechanisms are introduced into the attention framework. On the one hand, the dynamic attention score can be obtained from the scalar features $f_{ij}^{(0)}$. On the other hand, an explicit physical constraint is incorporated as the static attention score. The expanded radial features $\mathbf{R}(r_{ij})$ constructed from Gaussian radial basis functions are projected into learnable filter representations $\mathbf{F}_{ij}$ by the learnable matrix $\mathbf{W}_F$, and the atom-pair types $p_{ij} = z_j z_{\text{max}} + z_i$ encoded by one-hot encoding are projected into the representations of the pair-dependent coefficient $\mathbf{C}_{ij}$ and shift $\mathbf{D}_{ij}$ by the learnable matrices $\mathbf{W}_{C}$ and $\mathbf{W}_{D}$, respectively
\begin{align}
    \mathbf{F}_{ij} &= \mathbf{W}_F \mathbf{R}(r_{ij}), \\
    \mathbf{C}_{ij} &= \mathbf{W}_C \mathrm{OneHot}(p_{ij}), \\ 
    \mathbf{D}_{ij} &= \mathbf{W}_D \mathrm{OneHot}(p_{ij}).
\end{align}
Formally, the total attention score $s_{ij}$ between atoms $i$ and $j$ based on these two components is given by
\begin{equation}
    s_{ij} = \underbrace{ \mathbf{W}_f \, \psi_L (\mathrm{LayerNorm}(f_{ij}^{(0)})) }_{\text{Dynamic score}}
+ \underbrace{\langle \mathbf{C}_{ij} \odot  \mathbf{F}_{ij} + \mathbf{D}_{ij}, \mathbf{1} \rangle}_{\text{Static score}},
\end{equation}
where $\psi_L$ is the leaky ReLU activation, and the operator $\odot$ denotes the Hadamard product. The normalized attention weight $\alpha_{ij}$ is then computed using the Softmax function
\begin{equation} \label{weight}
    \alpha_{ij} = 
\frac{\exp(s_{ij})}
{\displaystyle\sum_{j' \in \mathcal{N}(i)} \exp(s_{ij'})},
\end{equation}
and the output embedding $\mathbf{x}_i^{(A)}$ of atom $i$ after the block of atom-pair-graph attention is updated as
\begin{equation} \label{x_A}
    \mathbf{x}_i^{(A)} = \mathbf{W}_A \sum_{j \in \mathcal{N}(i)} \alpha_{ij} \, \mathbf{v}_{ij}.
\end{equation}

\subsubsection{Scalar recurrent block}
Subsequently, a scalar recurrent block composed of the invariant fusion and gated recurrent unit (GRU)\cite{cho2014learning} is introduced to learn the harmonic and anharmonic embeddings of materials (Fig.~\ref{Fig1}d). The invariant fusion aims to scalarize the corresponding equivariant tensor features by retaining the  $l=0$ component and taking the $\ell_2$-norm over all $(l, m)$ channels, thereby yielding invariant scalar features that preserve the information from high-order tensors while ensuring the invariance of the crystal structure. The constructed two types of invariant features are concatenated, followed by LayerNorm
\begin{equation} \label{invariant}
    \mathbf{h}^{(t)}_i = \mathrm{LayerNorm} \left(\mathbf{x}_{i,(0,0)}^{(t)} \,\Vert\, \sqrt{
  \sum_{l=0}^{L} \sum_{m=-l}^{l}
   \vert \mathbf{x}_{i,(l,m)}^{(t)}  \vert^2
} \right),                                                                
\end{equation}
where $\mathbf{h}^{(t, \xi)}_i$ denotes the resulting invariant scalar representation of atom $i$ produced by the $t$-th update block. The index $t$ specifies whether the block corresponds to the initial graph embedding ($G$) or atom-pair-aware graph attention ($A$). 

GRU is further employed to adaptively fuse the newly updated feature $\mathbf{h}^{(S, \xi)}_i$ in
branch $\xi$ (indicating the harmonic branch (harm) or anharmonic branch (anharm)) as follows
\begin{align} 
    \mathbf{h}_i^{(S, \mathrm{Harm})} &= \mathbf{W}_{SH} \mathrm{GRU}^{(\mathrm{Harm})}(\mathbf{h}_i^{(G)}, \mathbf{h}_i^{(A)}), \label{gru_harm}\\
    \mathbf{h}_i^{(S, \mathrm{Anharm})} &= \mathbf{W}_{SA} \mathrm{GRU}^{(\mathrm{Anharm})}(\mathbf{h}_i^{(G)}, \mathbf{h}_i^{(S, \mathrm{Harm})}). \label{gru_anharm}
\end{align}

\subsubsection{Multi-gate Mixture-of-Experts}
Fig.~\ref{Fig1}e illustrates the Multi-gate Mixture-of-Experts (MMoE) framework designed to explicitly model the intrinsically coupled interactions between harmonicity and anharmonicity in materials.
Given the input representations $\mathbf{h}^{(S, \xi)}_i$, two task-specific gating networks dynamically regulate the harmonic and anharmonic expert pathways, enabling the model to adaptively emphasize their respective contributions. The gating weight $g_{i, k}^{(\xi)}$ are computed as
\begin{equation}
    g_{i, k}^{(\xi)} = 
\frac{\exp(\mathbf{W}_{g, k}^{(\xi)} \mathbf{h}_i^{(S, \xi)})}
{\displaystyle\sum_{k' \in K} \exp(\mathbf{W}_{g, k'}^{(\xi)} \mathbf{h}_i^{(S, \xi)})},
\end{equation}
where $k' \in K$ indexes the experts. Notably, the harmonic and anharmonic gating weights share a common Mixture-of-Experts (MoE) backbone, allowing the networks to capture the intrinsic coupling between these two physical components. The resulting representations are then obtained through the weighted aggregation of expert outputs
\begin{equation} \label{expert}
    \mathbf{h}_i^{(M, \xi)} = \sum_{k=1}^{K} g_{i, k}^{(\xi)} \mathrm{Expert}_k(\mathbf{h}_i^{(S, \xi)}).
\end{equation}

By explicitly embedding physical insights into the networks and leveraging the dynamic routing capability of the MoE module, the framework can adaptively optimize the contributions of latent features for harmonicity and anharmonicity, effectively bridging domain knowledge with ML.

\subsubsection{Physics formulation}
Our aim is to improve the data efficiency and generalization ability of $\kappa_\mathrm{L}$ prediction by embedding fundamental physical principles directly into the model prior to the ML procedure. To this end, we begin with the classical Slack model under dominant Umklapp scattering,\cite{morelli2008intrinsically, slack1973nonmetallic} a widely used physics model for estimating the $\kappa_\mathrm{L}$ of materials
\begin{equation}\label{slack}
    \kappa_{\mathrm{L}}
= A\, \frac{\overline{M}\,\delta\,\Theta_D^{3}}{\gamma^{2}\,T\, N^{2/3}},
\end{equation}
where A is a coefficient related to the Grüneisen parameter $\gamma$, defined as $A = 2.43 \times 10^{-6} \times (1-0.514/\gamma+0.228/{\gamma^2})^{-1}$. $\overline{M}$, $\Theta_{D}^{3}$, $\delta$, $N$, and $T$ are the average atomic mass, Debye temperature, volume per atom, number of atoms in the unit cell, and temperature, respectively. 
Given the non-negligible contributions of optical phonons to thermal transport,\cite{li2021optical, zheng2022anharmonicity} we employ the traditional Debye temperature $\Theta_D$ rather than the acoustic-mode Debye temperature, which is defined as\cite{anderson1963simplified}
\begin{equation}\label{debye}
    \Theta_D
=\frac{h}{k_B}\left(\frac{3N}{4\pi V}\right)^{\!1/3} \overline{\upsilon},
\end{equation}
where $h$, $k_B$, $V$, and $\overline{\upsilon}$ are the Planck constant, Boltzmann constant, cell volume, and average sound velocity, respectively. By inserting Eq.~(\ref{debye}) into Eq.~(\ref{slack}) and taking the base-10 logarithm of both sides, the resulting formula can be expressed as
\begin{align}\label{decompose}
    \log_{10} \kappa_\mathrm{L} =& \underbrace{ \log_{10}  \overline{\upsilon}^3 }_{\text{Harmonicity}}
+ \underbrace{ \log_{10} \frac{A}{\gamma^2} }_{\text{Anharmonicity}} \notag\\
&+ \underbrace{ \log_{10} \left( \frac{3\, \overline{M}\, \delta\, h^3\, N^{1/3}}{4\pi\, k_B^3\, T\, V} \right) }_{\text{Constant}}.
\end{align}
In this manner, $\log_{10} \kappa_\mathrm{L}$ is explicitly decomposed into the harmonicity term governed by $\overline{\upsilon}$ and the anharmonicity term governed by $\gamma$. Among them, we can obtain $\overline{\upsilon}$ as given by\cite{jia2017lattice}
\begin{align}
    \overline{\upsilon} &= \left[\frac{1}{3}\!\left(\upsilon_\mathrm{L}^{-3}+2\,\upsilon_\mathrm{T}^{-3}\right)\right]^{-1/3},  \\
    \upsilon_\mathrm{L} &= \sqrt{\frac{B + \frac{4}{3}G}{\rho}}, \\
    \upsilon_\mathrm{T} &= \sqrt{\frac{G}{\rho}} \label{velocity},
\end{align}
where $\upsilon_\mathrm{L}$, $\upsilon_\mathrm{T}$, and $\rho$ are the longitudinal sound velocity, transverse sound velocity, and density of the material, respectively.

Following the harmonicity–anharmonicity decomposition of $\kappa_{\mathrm{L}}$, we introduce this physics formulation block to map the learned harmonic and anharmonic representations (Fig.~\ref{Fig1}f). Specifically, node-level embeddings $\mathbf{h}_i^{(M, \xi)}$ are first aggregated into graph-level representations $\mathbf{h}_{\mathcal{G}}^{(\xi)}$ of the entire graph $\mathcal{G}$ for each branch $\xi$ through a graph readout function
\begin{equation}\label{readout}
    \mathbf{h}_{\mathcal{G}}^{(\xi)} = \mathrm{Readout} (\{ \mathbf{h}_i^{(M, \xi)} \, \vert \, \mathbf{v}_i \in \mathcal{G}\}),
\end{equation}
where the readout function is implemented using Global Attention Pooling,\cite{li2016gated} which adaptively aggregates node features according to their learned importance. The resulting graph-level representations are then passed through dense layers to predict the harmonic and anharmonic properties. For the harmonic branch, the predicted bulk modulus $\hat{B}$ and shear modulus $\hat{G}$ are given by
\begin{align}
    \hat{B} &= \mathrm{Dense}^{(B)} (\mathbf{h}_{\mathcal{G}}^{(\mathrm{Harm})}), \\
    \hat{G} &= \mathrm{Dense}^{(G)} (\mathbf{h}_{\mathcal{G}}^{(\mathrm{Harm})}).
\end{align}
For the anharmonic branch, the predicted Grüneisen parameter $\hat{\gamma}$ is given by
\begin{equation}
    \hat{\gamma} = \mathrm{Dense}^{(\gamma)}(\mathbf{h}_{\mathcal{G}}^{(\mathrm{Anharm})}).
\end{equation}

Accordingly, by inserting the predicted values $\hat{B}$, $\hat{G}$, and $\hat{\gamma}$ into Eqs.~(\ref{decompose})--(\ref{velocity}), the predicted $\hat{\kappa}_\mathrm{L}$ can be expressed in simplified form
\begin{equation}\label{physics}
    \log_{10} \hat{\kappa}_{\mathrm{L}}(\hat{B},\hat{G},\hat{\gamma} \mid  \boldsymbol{\theta})
  = \underbrace{ f(\hat{B},\hat{G} \mid \boldsymbol{\theta}) }_{\text{Harmonicity}} + \underbrace{ g(\hat{\gamma} \mid \boldsymbol{\theta}) }_{\text{Anharmonicity}} + C,
\end{equation}
where $\boldsymbol{\theta}$ denotes the learnable parameters of the model, and $C$ is constant. Finally, we incorporate this physics formulation into the ML model, giving rise to the physics-aware Transformer for $\kappa_\mathrm{L}$, termed KappaFormer. In contrast to the traditional end-to-end training of $\kappa_\mathrm{L}$, such a decomposable architecture explicitly embeds intrinsic physical insights.

\subsection{Model training and performance}
\begin{figure*}[htbp]
\centering
\includegraphics[width=1\textwidth]{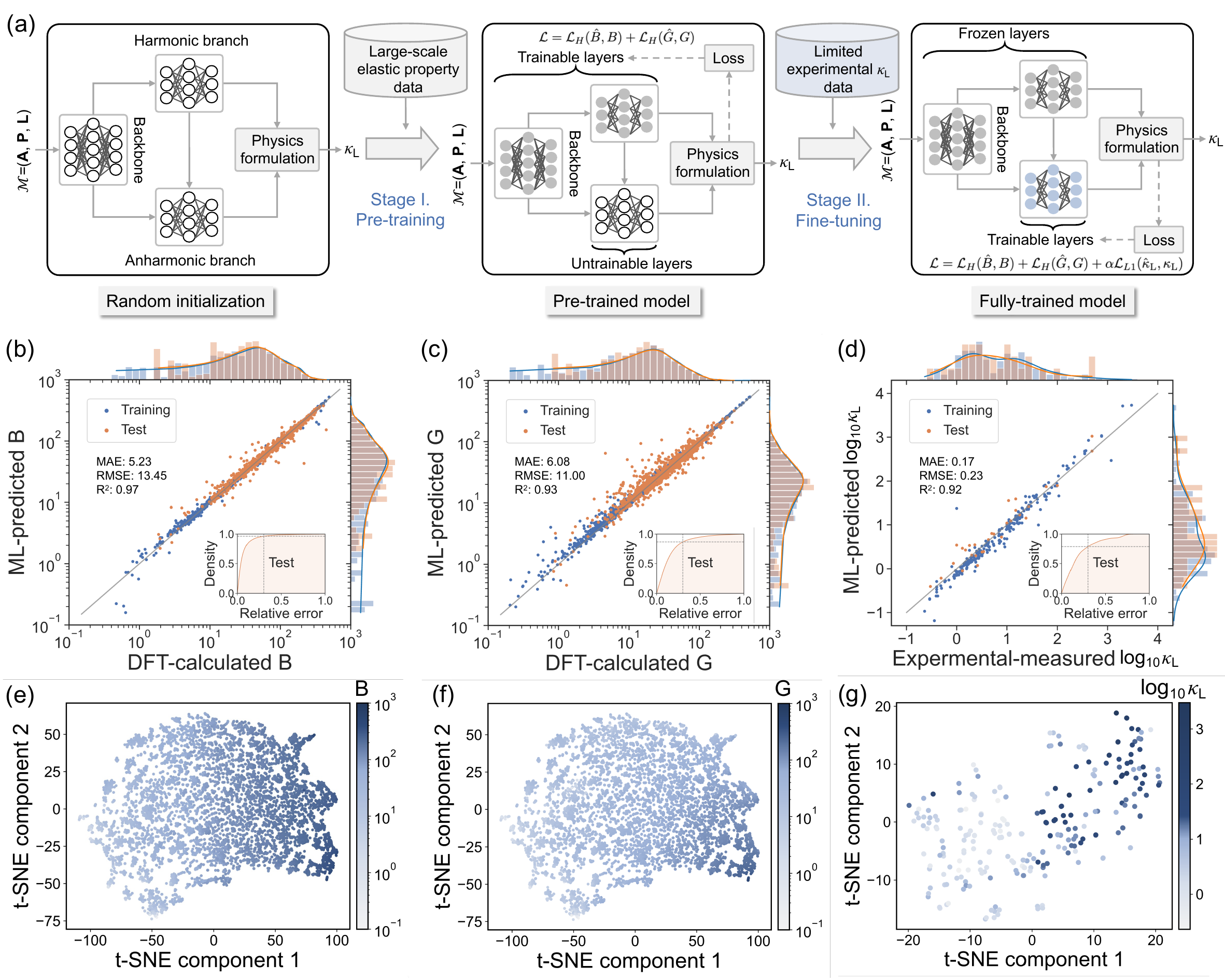}
\caption{\textbf{Model training and performance.} a) Dual-stage cross-domain transfer learning strategy to overcome data-scarce conditions. b)--d) ML-predicted values from KappaFormer compared with reference values for B, G, and $\kappa_\mathrm{L}$. The blue and orange data points denote the training and test set, respectively. The insets show the cumulative kernel density estimation (KDE) plots of the relative errors on the test sets. e)--g) t-SNE visualization of the hidden embeddings for B, G, and $\kappa_\mathrm{L}$ in the latent space. Each point indicates a distinct crystal structure.} 
\label{Fig2}
\end{figure*}

\subsubsection{Model training}
As indicated in Eq.~(\ref{physics}), $\kappa_\mathrm{L}$ has been formulated as linearly separable harmonic and anharmonic contributions. The elastic moduli $B$ and $G$, which govern the harmonicity term, can be accurately calculated using first-principles calculations, with the large-scale database of elastic properties already available. In contrast, the Grüneisen parameter $\gamma$, which governs the anharmonicity term, is considerably more challenging to determine. Accurate evaluation of $\gamma$ typically requires computationally demanding calculations, and its predictive accuracy remains limited. As a result, the uncertainty associated with anharmonicity represents a major source of error in the Slack model. Motivated by these observations, we propose an efficient dual-stage cross-domain transfer learning strategy that leverages large-scale elastic property data together with limited experimental $\kappa_\mathrm{L}$ data (Fig.~\ref{Fig2}a), as follows:

\textbf{Stage I: Pre-training on large-scale elastic property data.} The model is first randomly initialized and trained from scratch using large-scale elastic property data comprising over ten thousand entries retrieved from the MP database via its application programming interface (API). This pre-training stage enables the network to learn physically meaningful representations associated with harmonic properties, with the training objective optimized by the following total loss function
\begin{equation}
    \mathcal{L}_{\mathrm{pre}} = \mathcal{L}_H (\hat{B}, B) + \mathcal{L}_H (\hat{G}, G),
\end{equation}
where $\mathcal{L}_H$ denotes the Huber loss function.\cite{huber1992robust} After this stage, the model serves as a pre-trained model that captures transferable harmonic embeddings.

\textbf{Stage II: Fine-tuning on limited experimental $\kappa_\mathrm{L}$ data.} Starting from the pre-trained model, we perform fine-tuning on the experimental $\kappa_\mathrm{L}$ data from 302 distinct materials collected from the previous literature. To preserve the domain knowledge learned during pre-training, the harmonic branch is retained with frozen parameters, whereas the parameters of the anharmonic branch remain trainable. The experimental $\kappa_\mathrm{L}$ is incorporated as an additional training objective simultaneously, enabling the model to accurately resolve the anharmonicity, with the total loss function defined as
\begin{equation}
    \mathcal{L}_{\mathrm{ft}} = \mathcal{L}_H (\hat{B}, B) + \mathcal{L}_H (\hat{G}, G) + \alpha \mathcal{L}_{L1} (\hat{\kappa}_\mathrm{L}, \kappa_\mathrm{L}),
\end{equation}
where $\mathcal{L}_{L1}$ denotes the $L1$ loss function. After this stage, we obtain a fully-trained model capable of accurately predicting $\kappa_\mathrm{L}$ with physical interpretability. This dual-stage strategy facilitates effective knowledge transfer from elastic properties to $\kappa_\mathrm{L}$, thereby addressing the scarcity of experimental data for training.

\subsubsection{Model performance}
The comparisons of the KappaFormer predictions with reference values are shown in Fig.~\ref{Fig2}b--d. On the one hand, Fig.~\ref{Fig2}b and \ref{Fig2}c present the predicted bulk modulus $B$ and shear modulus $G$, respectively, both exhibiting strong agreement with the DFT-calculated values. On the test set, our model achieves a mean absolute error (MAE) of 5.23, a root mean squared error (RMSE) of 13.45, and the coefficient of determination ($R^2$) of 0.97 for $B$, while the prediction of $G$ yields a MAE of 6.08, a RMSE of 11.00, and an $R^2$ of 0.93, demonstrating state-of-the-art (SOTA) performance with respect to most representative baseline models (Table.~\ref{table.BG}),\cite{madani2025accelerating} namely SchNet,\cite{schutt2018schnet} CGCNN,\cite{xie2018crystal} MEGNet,\cite{chen2019graph} ALIGNN,\cite{choudhary2021atomistic} DeeperGATGNN,\cite{omee2022scalable} and Matformer.\cite{yan2022periodic} On the other hand, Fig.~\ref{Fig2}d compares the ML-predicted and experimental-measured $\log_{10} \kappa_\mathrm{L}$, demonstrating strong predictive performance with a MAE of 0.17, a RMSE of 0.23, and an $R^2$ of 0.92 on the test set. In addition, the marginal histograms further confirm the consistency between the predicted and reference distributions for $B$, $G$, and $\kappa_\mathrm{L}$, indicating that KappaFormer accurately captures both the overall statistics and value ranges of these properties. Simultaneously, 96\%, 87\%, and 79\% of the predictions on the test set for $B$, $G$, and $\kappa_\mathrm{L}$, respectively, have a relative error $\epsilon = |\hat{y} - y| / y$ below 30\% (see the insets of Fig.~\ref{Fig2}b--d), also demonstrating the strong predictive performance of KappaFormer across all three properties. 

To further understand the learned representations of materials, we visualize the latent embedding space using the t-distributed stochastic neighbor embedding (t-SNE)\cite{van2008visualizing} applied to the graph-level representations from the readout layer. Figures~\ref{Fig2}e and \ref{Fig2}f show the t-SNE embeddings colored by $B$ and $G$, respectively. A clear gradient in color is observed across the embedding space, indicating that the learned representations encode meaningful information related to the harmonic properties of materials. As for ${\kappa}_\mathrm{L}$, the embeddings used for the t-SNE visualization are obtained by concatenating the representations from the harmonic and anharmonic branches. Despite the limited size of the ${\kappa}_\mathrm{L}$ data, a consistent color gradient can still be observed in the latent space, indicating that the learned representations capture the underlying factors governing ${\kappa}_\mathrm{L}$.


\begin{table}[!t]
    \vspace{-3pt}
    \renewcommand{\arraystretch}{1.25}
     \tabcolsep=0.2cm
    \centering
    \caption{MAE of our model in comparison to the representative baseline models,\cite{madani2025accelerating} namely SchNet,\cite{schutt2018schnet} CGCNN,\cite{xie2018crystal} MEGNet,\cite{chen2019graph} ALIGNN,\cite{choudhary2021atomistic} DeeperGATGNN,\cite{omee2022scalable} and Matformer.\cite{yan2022periodic}}
    \label{table.BG}
    \begin{tabular}{ccc} 
        \toprule
        \textbf{Model}& \textbf{MAE ($\log_{10}B$)}& \textbf{MAE ($\log_{10}G$)}\\
        \midrule
        \textbf{KappaFormer} & \textbf{0.031} & \textbf{0.068} \\
        SchNet& 0.079 & 0.091  \\
        CGCNN& 0.073 & 0.085 \\
        MEGNet& 0.060 & 0.080 \\
        ALIGNN& 0.062 & 0.078 \\
        DeeperGATGNN& 0.069 & 0.093 \\
        Matformer& 0.059 & 0.076 \\
        \bottomrule
    \end{tabular}
\end{table}

\subsection{High-throughput material discovery}
\begin{figure*}[htbp]
\centering
\includegraphics[width=1\textwidth]{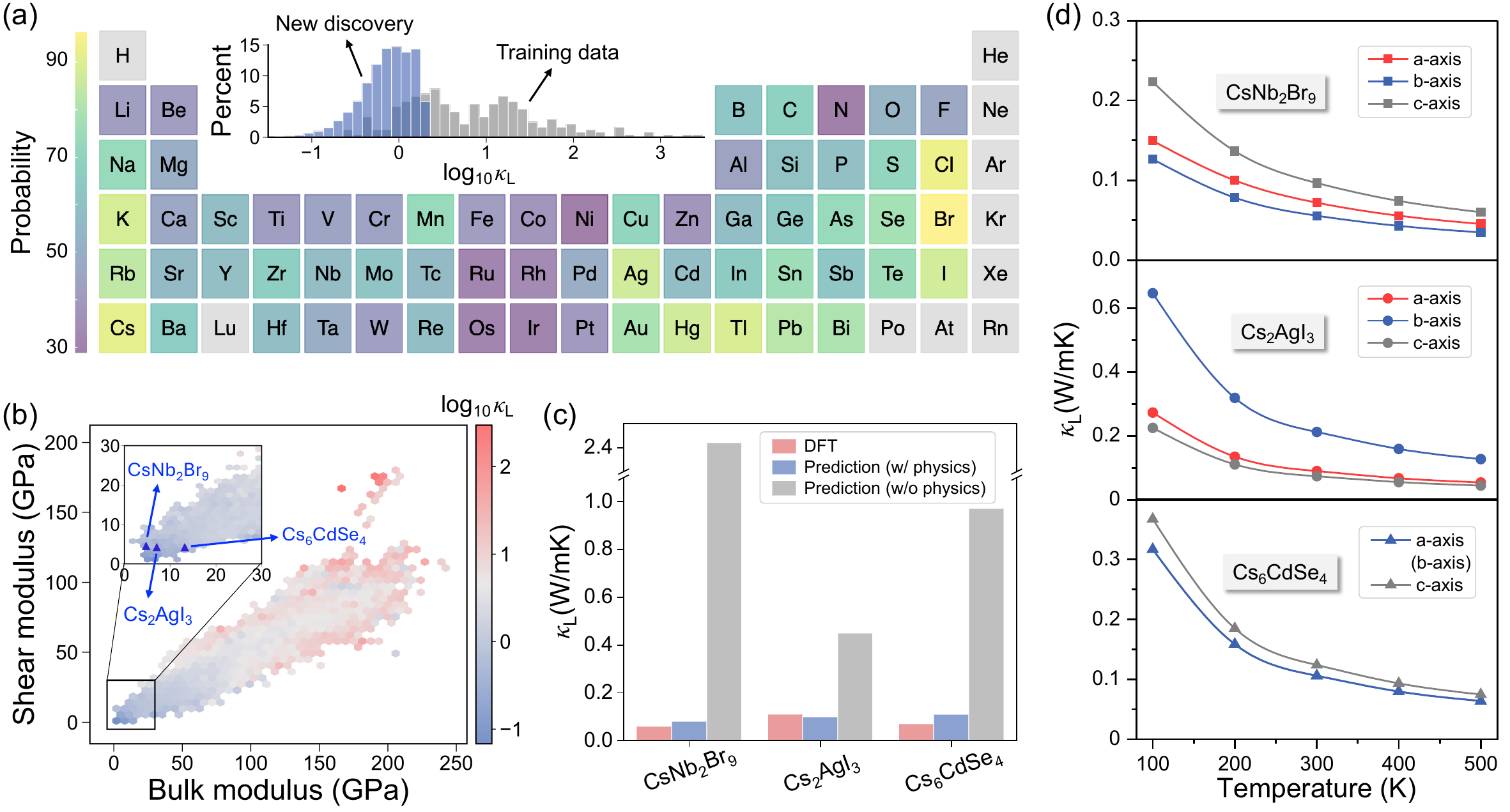}
\caption{\textbf{High-throughput material discovery.} a) Periodic table trends of elements associated with low $\kappa_\mathrm{L}$ materials, with the colour bar indicating the percentage probability of each element. Inset shows the histogram distribution of $\kappa_\mathrm{L}$ for the training data and newly discoverd materials with low $\kappa_\mathrm{L}$, where each dataset is normalized independently to sum to 100 percent. b) Scatter distribution of predicted \textit{B} and \textit{G} in the prediction dataset, colored by $\log_{10} \kappa_\mathrm{L}$. Inset highlights the low-modulus region and representative discovered materials with ultralow $\kappa_\mathrm{L}$, i.e., CsNb$_2$Br$_9$, Cs$_2$AgI$_3$, and Cs$_6$CdSe$_4$. c) Bar charts comparing DFT results with model predictions with (w/) and without (w/o) physics for the three materials. d) Temperature dependence of the DFT-calculated $\kappa_\mathrm{L}$ along different axes for the three materials.}
\label{Fig3}
\end{figure*}

\begin{figure*}[htbp]
\centering
\includegraphics[width=1\textwidth]{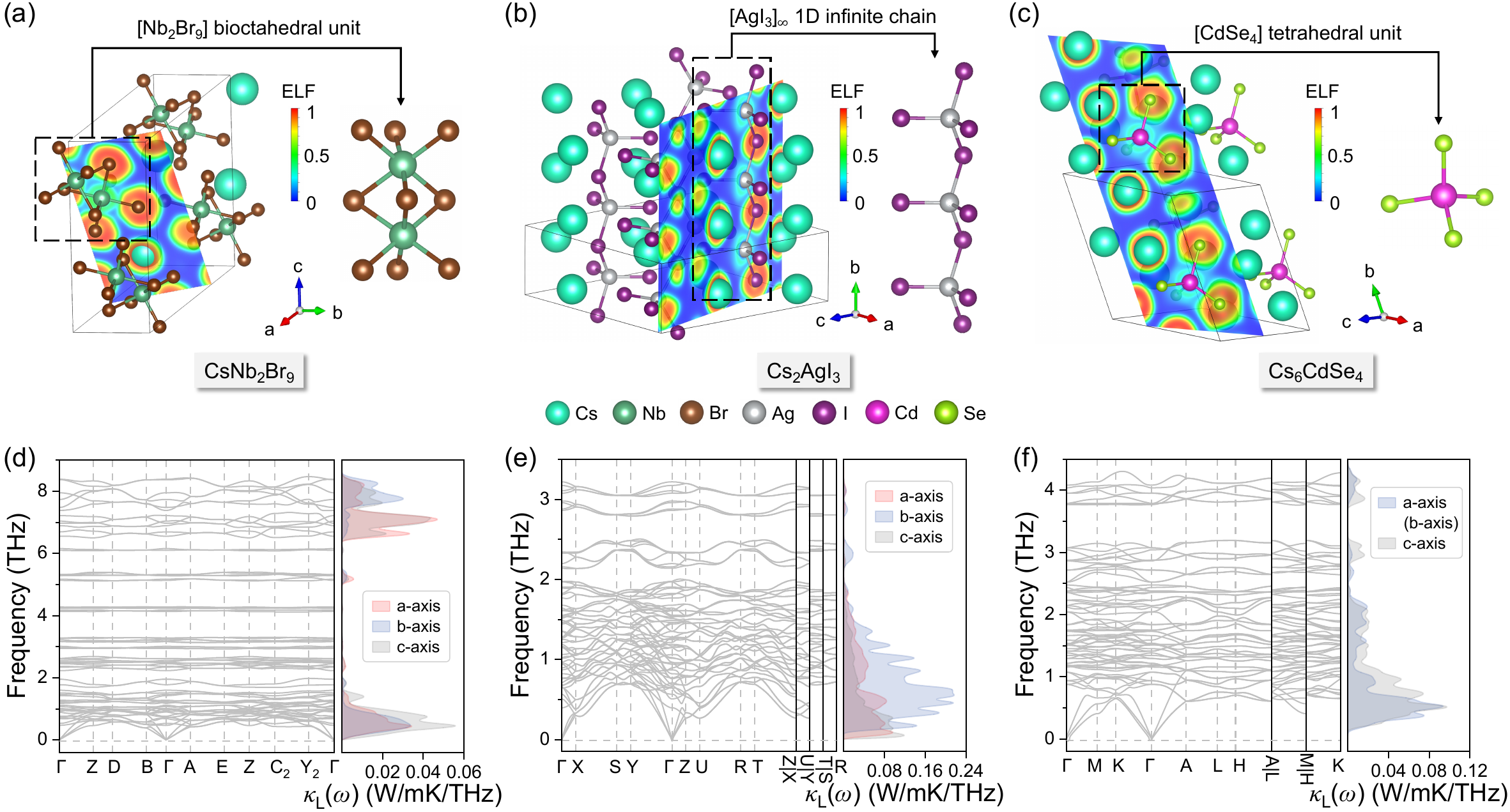}
\caption{\textbf{Structure characteristics and phonon thermal transport properties of the discovered three materials with ultralow $\kappa_\mathrm{L}$.} Crystal structures and the projected 2D ELF diagrams of a) CsNb$_2$Br$_9$, b) Cs$_2$AgI$_3$, and c) Cs$_6$CdSe$_4$. Phonon dispersion (left panels) and spectral $\kappa_\mathrm{L}(\omega)$ (right panels) of d) CsNb$_2$Br$_9$, e) Cs$_2$AgI$_3$, and f) Cs$_6$CdSe$_4$.}
\label{Fig4}
\end{figure*}

\subsubsection{Statistical analysis}
We next apply the fully-trained KappaFormer to the MP database to identify semiconductors with ultralow $\kappa_\mathrm{L}$, starting with the preliminary high-throughput screening of all MP entries. Thermodynamic stability is first assessed using the energy above the convex hull ($E_{\mathrm{hull}}$), with compounds satisfying $E_{\mathrm{hull}}$ $\leq$ 0.05 eV/atom retained as potentially synthesizable candidates. The band gap ($E_\mathrm{g}$) retrieved from the MP database is further restricted to 0.1--3 eV to ensure semiconducting behavior with reasonable electrical conductivity. Additionally, materials containing hydrogen, lanthanides, or actinides are excluded, yielding a prediction dataset of 24993 three-dimensional (3D) crystal structures. The $\kappa_\mathrm{L}$ values of these materials are ultimately predicted using our fully-trained model, thereby enabling efficient high-throughput material discovery.
 
Based on the predicted $\kappa_\mathrm{L}$ for tens of thousands of materials, we calculate the percentage probability of each element, resulting in periodic table trends as an initial guideline for designing materials with low $\kappa_\mathrm{L}$ (Fig.~\ref{Fig3}a). The percentage probability $p_m$ is defined as $p_m = N_m^{\mathrm{low}} / N_m \times 100\%$, where $N_m$ denotes the number of materials containing element $m$, and $N_m^{\mathrm{low}}$ is the number of those materials with $\kappa_\mathrm{L}$ $\leq$ 2 W/mK. This analysis reveals significant enrichment of halogens (Br, Cl, and I), alkali metals (Cs, K, and Rb), followed by several heavy metallic elements like Tl, Pb, Bi, Hg, etc., indicating that materials containing these elements are highly likely to possess low $\kappa_\mathrm{L}$. Such a trend is consistent with the previously discovered materials formed by combinations of these elements, including Cs$_2$AgBiBr$_6$,\cite{zheng2024unravelling} Cs$_3$Bi$_2$Br$_9$,\cite{li2024phonon} Cs$_3$Bi$_2$I$_6$Cl$_3$,\cite{acharyya2022glassy} K$_2$Ag$_4$Se$_3$,\cite{li2025long} and Tl$_3$VSe$_4$,\cite{mukhopadhyay2018two} and so forth.
The inset of Fig.~\ref{Fig3}a presents the histogram distribution of $\kappa_\mathrm{L}$ for the training data and the newly discovered materials with $\kappa_\mathrm{L}$ $\leq$ 2 W/mK (more than ten thousand), where the identified materials are mainly located towards the left side of the training data. Quantitatively, the average $\kappa_\mathrm{L}$ of these materials is 0.89 W/mK, whereas the training data exhibit a mean value of 56.53 W/mK, representing the expected level for randomly selected materials. These results demonstrate the effectiveness of the fully trained KappaFormer in identifying materials with low $\kappa_\mathrm{L}$.

Figure~\ref{Fig3}b illustrates the scatter distribution of predicted \textit{B} and \textit{G} in the prediction dataset, colored by $\log_{10} \kappa_\mathrm{L}$, which reveals a clear positive correlation between the two elastic moduli. Materials with lower $\kappa_\mathrm{L}$ tend to cluster in the region of relatively small elastic moduli, in agreement with the general understanding that mechanically softer materials often exhibit reduced $\kappa_\mathrm{L}$. The inset of Fig.~\ref{Fig3}b highlights three representative discovered materials labeled in the low-modulus region, i.e., CsNb$_2$Br$_9$, Cs$_2$AgI$_3$, and Cs$_6$CdSe$_4$, which are selected from thermodynamically stable materials with the lowest predicted $\kappa_\mathrm{L}$ containing fewer than 25 atoms per unit cell that remain previously unreported. These materials are further validated by DFT calculations.

\subsubsection{DFT validation}
According to DFT-derived IFCs and phonon Boltzmann transport theory, we obtain more accurate $\kappa_\mathrm{L}$ values for CsNb$_2$Br$_9$, Cs$_2$AgI$_3$, and Cs$_6$CdSe$_4$, followed by a bar-chart comparison with the model predictions, as shown in Fig.~\ref{Fig3}c. Simultaneously, we also train an end-to-end baseline without the physics formulation as an ablation model for comparison, which yields a MAE of 0.24 on the test set. The predictions incorporating physics show excellent agreement with the DFT results, whereas the model without physics significantly overestimates $\kappa_\mathrm{L}$. This comparison highlights that incorporating physical insights into ML not only achieves high predictive accuracy but also enhances the generalization ability of the model in predicting $\kappa_\mathrm{L}$. Fig.~\ref{Fig3}d depicts the DFT-calculated $\kappa_\mathrm{L}$ as a function of temperature ($T$) ranging from 100 to 500 K along different axes for the discussed three materials. The intrinsic $\kappa_\mathrm{L}$ values show obvious anisotropy and gradually decrease with increasing temperature, approximately following a $T^{-1}$ manner, consistent with the characteristic behavior governed by Umklapp phonon scattering. The rise in temperature results in a higher equilibrium phonon population, thereby enhancing phonon–phonon scattering and promoting the Umklapp process, which shortens the phonon lifetime and suppresses $\kappa_\mathrm{L}$.

CsNb$_2$Br$_9$, Cs$_2$AgI$_3$, and Cs$_6$CdSe$_4$ crystallize in the monoclinic $P2/c$, orthorhombic $Pnma$, and hexagonal $P6_3mc$ space groups, respectively, with the fully relaxed crystallographic parameters listed in {\color{blue}{Tables S1--S3}} and crystal structures shown in Fig.~\ref{Fig4}a--c. For CsNb$_2$Br$_9$, two neighboring [NbBr$_6$] halide octahedra connect via face-sharing to form [Nb$_2$Br$_9$] bioctahedral units oriented along the \textit{a}–\textit{c} plane. For Cs$_2$AgI$_3$, the structure features [AgI$_3$]$_\infty$ one-dimentional (1D) infinite chains formed by corner-sharing [AgI$_4$] tetrahedra runing along \textit{b}-axis. In the case of Cs$_6$CdSe$_4$, Cd atom is tetrahedrally coordinated by four Se atoms, forming isloated [CdSe$_4$] tetrahedra. In all three compounds, the Cs$^+$ cations fill the spaces between these structural motifs, stabilizing the extended lattice.
To characterize bonding in these materials, we employ the electron localization function (ELF), which quantifies the spatial localization of electrons with values ranging from 0 to 1 (Fig.~\ref{Fig4}a--c). The projected 2D ELF diagrams show that Cs atoms form highly ionic bonds with surrounding anions in all cases, as evidenced by the absence of electron localization in the Cs–X (X = Br, I, and Se) bonding regions. In contrast, the Nb–Br, Ag–I, and Cd–Se bonds exhibit varying degrees of electron localization. A polar covalent bond between Nb and Br (ELF $\approx$ 0.5) is observed, indicating a stronger covalent contribution, whereas the Ag–I and Cd–Se interactions are dominated by ionic character. Simultaneously, the phonon dispersion relations along high-symmetry paths in the Brillouin zone exhibit no imaginary modes (left panels in Fig.~\ref{Fig4}d--f), confirming the dynamical stability of these discovered materials, which supports their experimental feasibility. Further analysis of the spectral $\kappa_{\mathrm{L}}(\omega)$ elucidates the frequency-resolved contributions to heat transport in these compounds (right panels in Fig.~\ref{Fig4}d--f). For CsNb$_2$Br$_9$, both low-frequency acoustic phonons and high-frequency optical modes make substantial contributions to $\kappa_{\mathrm{L}}$, indicating a non-negligible participation of optical branches in heat transport. In contrast, for Cs$_2$AgI$_3$ and Cs$_6$CdSe$_4$, $\kappa_{\mathrm{L}}$ is predominantly governed by low-frequency phonon modes, with high-frequency optical phonons contributing only marginally.

\subsection{Physics interpretability of KappaFormer}
\begin{figure}[htbp]
\centering
\includegraphics[width=0.5\textwidth]{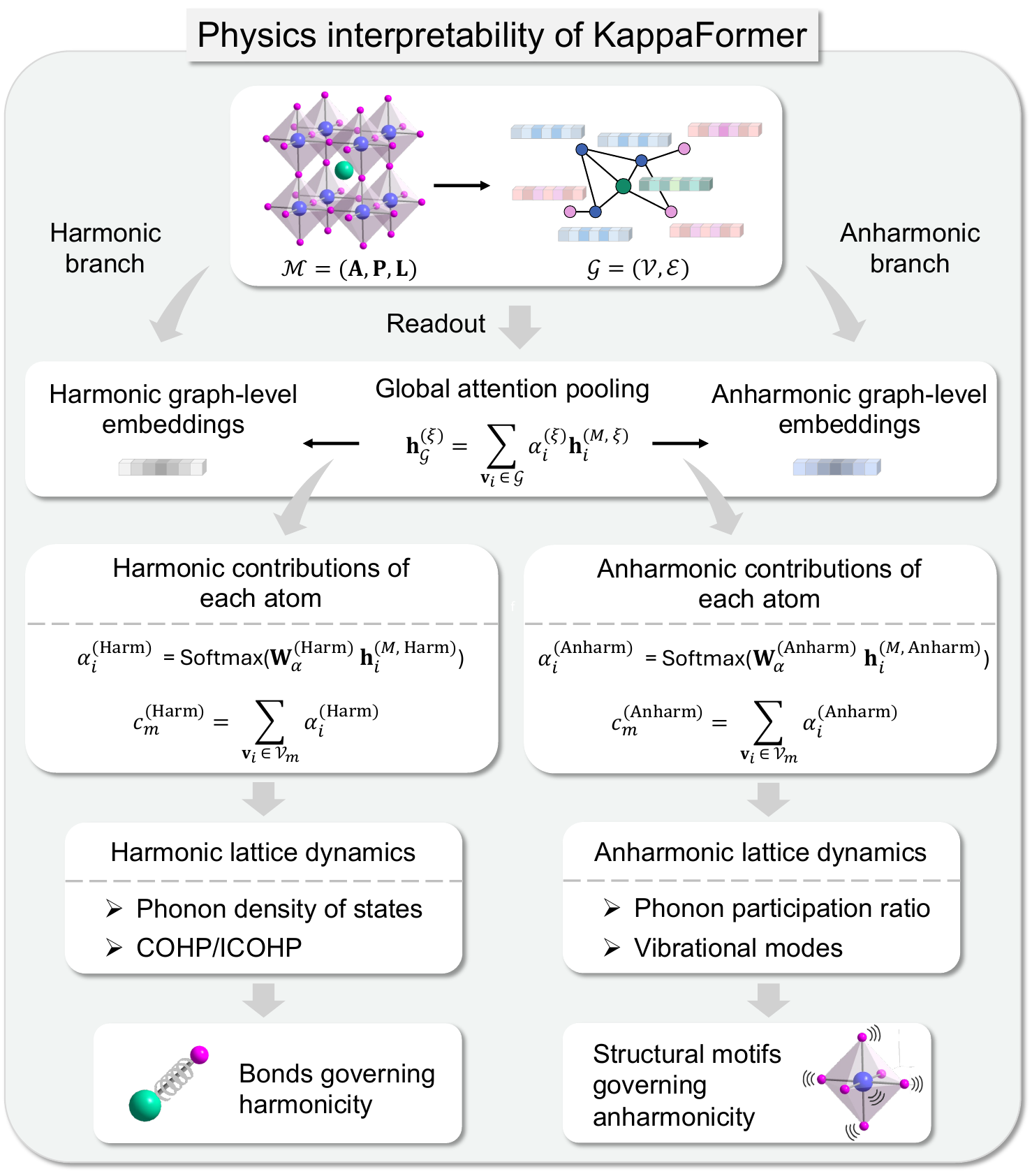}
\caption{\textbf{Physics interpretability of KappaFormer.} The whole framework including deriving atom-resolved contributions from KappaFormer and elucidating phonon thermal transport mechanisms from DFT calculations.}
\label{Fig5}
\end{figure}

\begin{figure*}[htbp]
\centering
\includegraphics[width=1\textwidth]{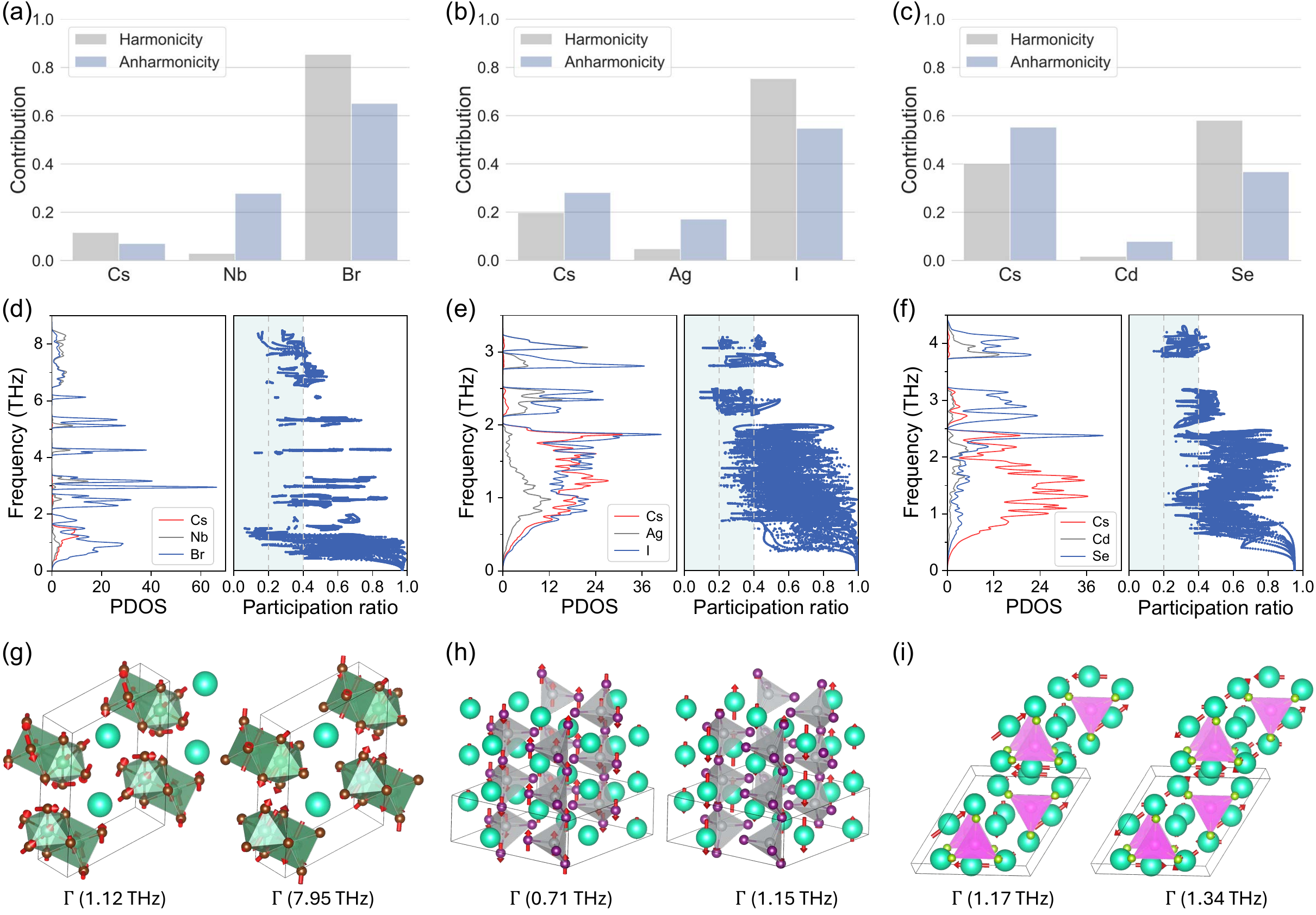}
\caption{\textbf{Harmonic and anharmonic interpretable analysis of the discovered three materials with ultralow $\kappa_\mathrm{L}$.} Atom-resolved contributions to harmonicity and anharmonicity of a) CsNb$_2$Br$_9$, b) Cs$_2$AgI$_3$, and c) Cs$_6$CdSe$_4$. PDOS and PPR of d) CsNb$_2$Br$_9$, e) Cs$_2$AgI$_3$, and f) Cs$_6$CdSe$_4$. Visualization of the localized vibrational modes of g) CsNb$_2$Br$_9$, h) Cs$_2$AgI$_3$, and i) Cs$_6$CdSe$_4$ at the $\Gamma$ point.}
\label{Fig6}
\end{figure*}

\subsubsection{Interpretable framework}
Despite the strong predictive performance in predicting $\kappa_\mathrm{L}$, the underlying physical mechanisms of phonon thermal transport remain difficult to resolve from the model. To uncover these mechanisms, we introduce an interpretable framework of KappaFormer that bridges the learned embeddings to the harmonic and anharmonic behavior of materials as shown in Fig.~\ref{Fig5}.

Starting from the crystal structure, node-level embeddings $\mathbf{h}_i^{(M, \xi)}$ are updated and subsequently aggregated into a graph-level representation $\mathbf{h}_{\mathcal{G}}^{(\xi)}$ via a readout function. Regarding Eq.~\ref{readout}, this readout function is implemented using Global Attention Pooling, which assigns the contributions of each node for the graph $\mathcal{G}$ in each branch $\xi$
\begin{equation}
\mathbf{h}_{\mathcal{G}}^{(\xi)}
=\sum_{\mathbf{v}_i \in \mathcal{G}}
\alpha_i^{(\xi)} \mathbf{h}_i^{(M,\xi)},
\end{equation}
where the attention weight $\alpha_i^{(\xi)}$ of atom \textit{i} is defined as
\begin{equation}
\alpha_i^{(\xi)}
= \frac{
\exp ( \mathbf{W}^{(\xi)}_{\alpha} \mathbf{h}_i^{(M,\xi)} )
}{\displaystyle
\sum_{\mathbf{v}_j \in \mathcal{G}}
\exp ( \mathbf{W}^{(\xi)}_{\alpha} \mathbf{h}_j^{(M,\xi)} )
},
\end{equation}
where $\mathbf{W}^{(\xi)}_{\alpha}$ is the learnable matrix. To obtain atom-resolved contributions, we further aggregate the attention weights over atoms belonging to the same chemical species. Let
\begin{equation}
\mathcal{V}_m = \{\mathbf{v}_i \in \mathcal{V} \mid Z_i = m \},
\end{equation}
where $\mathcal{V}$ denotes the set of atoms in graph $\mathcal{G}$, $\mathcal{V}_m$ denotes the subset of atoms in $\mathcal{V}$ corresponding to the chemical element $m$, and $Z_i$ represents the chemical species of atom $i$. The contribution of element $m$ in branch $\xi$ is then defined as
\begin{equation}
c_m^{(\xi)} =
\sum_{\mathbf{v}_i \in \mathcal{V}_m} \alpha_i^{(\xi)}.
\end{equation}
As such, we can obtain the harmonic and anharmonic contributions of each atom in a crystal structure.

To bridge the atom-resolved contributions derived from KappaFormer with physical quantities, DFT-level analyzes are performed to elucidate the roles of harmonicity and anharmonicity in phonon thermal transport. On the one hand, we analyze the phonon density of states (PDOS) to quantify the contributions of each atom to the phonon spectrum. By further calculating the corresponding crystal orbital Hamilton population (COHP) and its integral ($\mathrm{ICOHP}=\int_{-\infty}^{E_\mathrm{F}} \mathrm{COHP}(E)\mathrm{d}E$) at the Fermi level $E_\mathrm{F}$ between atomic pairs to identify bond strengths,\cite{dronskowski1993crystal} we are able to reveal the bonds that govern harmonicity.
On the other hand, we introduce the phonon participation ratio (PPR), which serves as a widely used metric to characterize the extent of localization of the phonon modes. The PPR for a phonon mode with frequency $\omega_q$ is defined as\cite{pailhes2014localization, tadano2015impact}
\begin{equation}
P(\omega_q) =
\frac{
\left( \sum_{i=1}^{N} \left| e_i(\omega_q) \right|^2 / {M_i} \right)^2
}{
N \sum_{i=1}^{N} \left| e_i(\omega_q) \right|^4 / {M_i^2}
},
\end{equation}
where $e_i(\omega_q)$ denotes the eigenvector of the phonon mode at wave vector $q$ with frequency $\omega$, $M_i$ is the mass of the $i$th atom, and $N$ is the number of atoms in the unit cell. Notably, $P(\omega_q)$ ranges from 0 to 1, with values approaching 1 indicating delocalized (propagating) phonon modes involving nearly all atoms in the unit cell, whereas lower values correspond to strongly localized modes dominated by only a small subset of atoms.\cite{beltukov2013ioffe, pailhes2014localization} This enables direct visualization of the corresponding strongly localized vibrational modes, thereby revealing the structural motifs governing anharmonicity.

\subsubsection{Harmonic interpretability}
As shown in Fig.~\ref{Fig6}a--c, the harmonic atom-resolved contributions are predominantly governed by the halogen/chalcogen atoms in CsNb$_2$Br$_9$, Cs$_2$AgI$_3$, and Cs$_6$CdSe$_4$, with Br, I, and Se contributing approximately 0.85, 0.75, and 0.58, respectively, showing a gradually decreasing trend. The contribution from Cs increases from 0.12 to 0.20 and further to 0.40, whereas Nb, Ag, and Cd exhibit negligible contributions to the harmonicity. These results reflect a shift in harmonic contributions from the anionic atoms to the Cs atoms.

This trend can be rationalized by the PDOS results (left panels in Fig.~\ref{Fig6}d--f), which show that the framework anions (Br, I, and Se) consistently make substantial contributions to the harmonic lattice dynamics across all three compounds, while the participation of Cs atoms in the low-frequency region progressively increases, indicating an enhanced role of Cs-related phonon modes in the harmonic response. Further insights are provided by the COHP analysis, which shows that the framework bonding strength follows the order Nb--Br $>$ Ag--I $>$ Cd--Se, suggesting a more pronounced covalent character for Nb--Br ({\color{blue}{Fig.~S1a}}). Meanwhile, all Cs--X (X = Br, I, and Se) interactions exhibit very small ICOHP values, indicative of intrinsically weak ionic bonding ({\color{blue}{Fig.~S1b}}). Such weak Cs--X bonds lead to softened harmonic IFCs, which favor low phonon group velocities involving Cs atoms and reduce the overall stiffness of the lattice. This can be rationalized in terms of the dispersion relation for the frequency, just as $\omega=2\sqrt{{\beta}/M}\left|\sin{qa}/2\right|$ in a one-dimensional crystal lattice, where $\beta$ and $M$ are the bond force constant and the atomic mass, respectively. Weak ionic bonding of Cs--X can reduce $\beta$, leading to lower phonon frequencies and group velocities ({\color{blue}{Fig.~S2}}). This suppression of harmonic phonon transport ultimately contributes to the ultralow $\kappa_\mathrm{L}$. 

\subsubsection{Anharmonic interpretability}
For the anharmonicity of CsNb$_2$Br$_9$, Cs$_2$AgI$_3$, and Cs$_6$CdSe$_4$, in addition to the dominant contributions from the halogen/chalcogen atoms, Nb, Ag, and Cd also make non-negligible contributions, which are significantly larger than their harmonic terms, highlighting their pronounced role in the anharmonic lattice response, with contributions of approximately 0.28, 0.17, and 0.08, respectively. Among them, Nb shows the most substantial enhancement, followed by Ag, while Cd also exhibits a noticeable increase, whereas the contribution from Cs increases markedly from 0.07 to 0.28 and further to 0.55. These results indicate that the anharmonicity in CsNb$_2$Br$_9$ is primarily governed by Nb and Br, while in Cs$_2$AgI$_3$, both Cs and the framework atoms (Ag and I) play important roles. In contrast, Cs becomes the dominant contributor to the anharmonicity in Cs$_6$CdSe$_4$.

The frequency-dependent PPR reveals a common feature among CsNb$_2$Br$_9$, Cs$_2$AgI$_3$, and Cs$_6$CdSe$_4$ as shown in the right panels of Fig.~\ref{Fig6}d--f, namely the presence of relatively low PPR values around $\sim$0.4, indicative of substantial phonon localization. Notably, CsNb$_2$Br$_9$ exhibits even smaller PPR values, reaching as low as $\sim$0.2 near both the low-frequency region and the high-frequency optical region, indicating stronger localization behavior than in Cs$_2$AgI$_3$ and Cs$_6$CdSe$_4$, while these frequency ranges also contribute significantly to $\kappa_{\mathrm{L}}$ as discussed above. Further, we visualize the atomic vibration eigenvectors of these phonon modes with low PPR values in Fig.~\ref{Fig6}g--i and {\color{blue}{Fig.~S3}}, which are also characterized by short phonon lifetimes ({\color{blue}{Fig.~S4}}). In CsNb$_2$Br$_9$, the modes at the $\Gamma$ point around 1.12 THz and 7.95 THz are primarily governed by the collective rotational motions of the [Nb$_2$Br$_9$] bioctahedral units, accompanied by minor contributions from the rattling vibrations of Cs atoms around 1.21 THz. In Cs$_2$AgI$_3$, the low-frequency modes around 0.71 THz and 1.15 THz are dominated by the coupled collective motions of [AgI$_3$]$_\infty$ 1D infinite chains and the rattling vibrations of Cs atoms, with comparable contributions from both components, where the chain vibrations feature out-of-phase atomic displacements propagating along the \textit{b}-axis. The collective motions of [AgI$_3$]$_\infty$ 1D chains can also be observed around 2.34~THz, where the PPR decreases to $\sim$0.2, yet these modes contribute little to $\kappa_{\mathrm{L}}$. As for Cs$_6$CdSe$_4$, the low-frequency acoustic modes with the lowest PPR values are primarily governed by the rattling vibrations of Cs atoms at the $\Gamma$ point around 1.17 THz and 1.34 THz, while the highly localized high-frequency modes ($\sim$0.2) around 3.79 THz with slight contributions to $\kappa_\mathrm{L}$ are associated with the [CdSe$_4$] tetrahedral units. As a result, these vibrational modes associated with specific structural motifs substantially reduce phonon lifetimes and induce strong lattice anharmonicity, which suppress heat transport and leads to intrinsically ultralow $\kappa_\mathrm{L}$ of these materials.

\section{Discussion}  
In this work, we develop KappaFormer as a physics-aware Transformer for predicting $\kappa_\mathrm{L}$ by explicitly embedding harmonic–anharmonic decomposition into a graph-based attention architecture. This design facilitates effective knowledge transfer from large-scale elastic data to data-scarce $\kappa_\mathrm{L}$ data, achieving both high predictive accuracy and robust generalization. The fully trained model  enables large-scale screening across extensive material spaces, identifying candidates with intrinsically ultralow $\kappa_{\mathrm{L}}$, among which CsNb$_2$Br$_9$, Cs$_2$AgI$_3$, and Cs$_6$CdSe$_4$ are validated via phonon BTE. Beyond prediction, we introduce a physics-interpretable framework that bridges atom-resolved contributions from KappaFormer with physical quantities derived from DFT analysis, revealing the bonds that govern harmonicity and the structural motifs that govern anharmonicity. Collectively, our findings suggest a general design principle for ultralow $\kappa_{\mathrm{L}}$ materials, namely the synergistic interplay between soft lattice frameworks and localized vibrational units such as rattling cations or flexible polyhedral motifs. Overall, this study demonstrates that integrating physical insights with advanced ML architectures provides a powerful and generalizable route for understanding and discovering materials with target properties.

\section{Methods}  
\subsection{$SO(2)$-tensor product}



To improve the efficiency of $SO(3)$-equivariant message passing, the Clebsch--Gordan tensor product (CGTP) can be reformulated in a local coordinate frame, leading to an induced $SO(2)$ structure.\cite{escn, griffiths2018introduction} Regarding Eq.~(\ref{so2}), the function $\mathcal{T}(\cdot)$ corresponds to this $SO(2)$-TP, which operates on pairs of modes $(m,-m)$ in a block-wise manner.
Starting from the standard $SO(3)$-equivariant message\cite{geiger2022e3nn, goodman2000representations} 
\begin{equation}
\mathbf{a}_{st}^{(l_o)}
=
\sum_{l_i,l_f}
\mathbf{x}_s^{(l_i)}
\otimes_{l_i,l_f}^{\,l_o}
\,\mathbf{h}_{l_i,l_f,l_o}\,
\mathbf{Y}^{(l_f)}(\hat{\mathbf r}_{st}),
\label{eq:so3_msg}
\end{equation}
where $l_i$, $l_f$, and $l_o$ denote the degrees of the input, filter, and output irreps, respectively. $\otimes^{\,l_o}_{l_i,l_f}$ is the CGTP, $s$ and $t$ denote two neighboring atoms, corresponding to the source and target nodes that define the edge direction $\mathbf{r}_{st}$. $\mathbf{Y}^{(l_f)}(\hat{\mathbf r}_{st})$ is the spherical harmonics basis function,\cite{Gasteiger2020Directional} where $\hat{\mathbf{r}}_{st} = \mathbf{r}_{st} / r_{st}$ is the unit vector.
An edge-dependent rotation $\mathbf R_{st}$ is introduced such that the edge direction $\hat{\mathbf r}_{st}$ is aligned with a fixed axis. The input feature is rotated accordingly,
\begin{equation}
\tilde{\mathbf x}_s^{(l_i)}=
\mathbf D^{(l_i)}(\mathbf R_{st})\mathbf x_s^{(l_i)}.
\label{eq:rot_feat}
\end{equation}
By equivariance, Eq.~\eqref{eq:so3_msg} can be evaluated in the local frame as
\begin{equation}
\mathbf a_{st}^{(l_o)}
=
\mathbf D^{(l_o)}(\mathbf R_{st}^{-1})
\sum_{l_i,l_f}
\tilde{\mathbf x}_s^{(l_i)}
\otimes_{l_i,l_f}^{\,l_o}
\,\mathbf{h}_{l_i,l_f,l_o}\,
\mathbf Y^{(l_f)}(\mathbf R_{st}\hat{\mathbf r}_{st}).
\label{eq:local_msg}
\end{equation}
In this aligned frame, the spherical harmonic becomes sparse as given by
\begin{equation}
\mathbf{Y}_m^{(l)}\!\left(\mathbf{R}_{st}\hat{\mathbf r}_{st}\right)
\propto
\delta_m^{(l)}
=
\begin{cases}
1, & m = 0,\\
0, & m \neq 0.
\end{cases}
\end{equation}
The summation over $l_f$ together with the CG coefficients can be absorbed into an effective $SO(2)$ filter,
\begin{equation}
\mathbf{\tilde h}_m^{(l_i,l_o)}
=
\sum_{l_f}
\mathbf{h}_{l_i,l_f,l_o}\,(\mathbf{c}_{l_i,l_f,l_o})_m,
\label{eq:eff_filter}
\end{equation}
leading to a decomposition into independent $SO(2)$ blocks. For $m_o>0$, the paired modes $(m,-m)$ are updated as
\begin{equation}
\begin{pmatrix}
(\mathbf w_{st,l_i}^{(l_o)})_{m_o}\\
(\mathbf w_{st,l_i}^{(l_o)})_{-m_o}
\end{pmatrix}
=
\begin{pmatrix}
\mathbf {\tilde h}_{m_o}^{(l^{\prime},l)} & -\mathbf{\tilde h}_{-m_o}^{(l^{\prime},l)}\\
\mathbf{\tilde h}_{-m_o}^{(l^{\prime},l)} & \mathbf{\tilde h}_{m_o}^{(l^{\prime},l)}
\end{pmatrix}
\cdot
\begin{pmatrix}
(\tilde{\mathbf x}_s^{(l_i)})_{m_o}\\
(\tilde{\mathbf x}_s^{(l_i)})_{-m_o}
\end{pmatrix},
\label{eq:so2_block_short}
\end{equation}
while the $m_o=0$ channel is updated independently,
\begin{equation}
(\mathbf w_{st,l_i}^{(l_o)})_0
=
\mathbf{\tilde h}_0^{(l_i,l_o)}(\tilde{\mathbf x}_s^{(l_i)})_0.
\label{eq:so2_m0_short}
\end{equation}
The final message is obtained by summing over input irreps and rotating back to the global frame,
\begin{equation}
\mathbf a_{st}^{(l_o)}
=
\mathbf D^{(l_o)}(\mathbf R_{st}^{-1})
\sum_{l_i}\mathbf w_{st,l_i}^{(l_o)}.
\label{eq:final_so2_short}
\end{equation}
This formulation preserves global $SO(3)$ equivariance while replacing the expensive $SO(3)$ CGTP with efficient local $SO(2)$ block convolutions.

\subsection{Gated recurrent unit}
For the GRU function\cite{cho2014learning} described in Eqs.~\ref{gru_harm} and \ref{gru_anharm}, given the initial graph embedding $\mathbf{h}^{(G)}_i$ and the updated embedding $\mathbf{h}^{(t)}_i$ (\textit{t} = \textit{A} or Harm), we here provide the explicit formulation
\begin{align}
\mathbf{z}^{(\xi)}_i &= \sigma(\mathbf{W}_z^{(\xi)} \mathbf{h}^{(t)}_i + \mathbf{U}_z^{(\xi)} \mathbf{h}^{(G)}_i + \mathbf{b}_z^{(\xi)}), \\
\mathbf{r}^{(\xi)}_i &= \sigma(\mathbf{W}_r^{(\xi)} \mathbf{h}^{(t)}_i + \mathbf{U}_r^{(\xi)} \mathbf{h}^{(G)}_i + \mathbf{b}_r^{(\xi)}), \\
\tilde{\mathbf{h}}^{(\xi)}_i &= \phi(\mathbf{W}_{\tilde{h}}^{(\xi)} \mathbf{h}^{(t)}_i + \mathbf{U}_{\tilde{h}}^{(\xi)} (\mathbf{r}^{(\xi)}_i \odot \mathbf{h}^{(G)}_i) + \mathbf{b}_{\tilde{h}}^{(\xi)}), \\
\mathbf{h}^{(\xi)}_i &= [(1 - \mathbf{z}^{(\xi)}_i) \odot \mathbf{h}^{(G)}_i + \mathbf{z}^{(\xi)}_i \odot \tilde{\mathbf{h}}^{(\xi)}_i],
\end{align}
where $\mathbf{z}^{(\xi)}_i$, $\mathbf{r}^{(\xi)}_i$, $\tilde{\mathbf{h}}^{(\xi)}_i$, and $\mathbf{h}^{(\xi)}_i$ are the update gate, reset gate, candidate hidden state, and fused hidden state of atom $i$ in branch $\xi$, respectively. $\sigma(\cdot)$ and $\phi(\cdot)$ are the sigmoid activation function and the tanh activation function, $\mathbf{W}$ and $\mathbf{U}$ are the learnable weights, and $\mathbf{b}$ is the learnable bias.

\subsection{Implementation details of KappaFormer}
The KappaFormer model was implemented using PyTorch,\cite{paszke2019pytorch} in which the training and inference processes were carried out on a single NVIDIA A800 Tensor Core GPU equipped with 80 GB of high-bandwidth memory. We optimized the model parameters by using the AdamW optimizer with standard hyperparameters. For the elastic property task, the model was trained with a batch size of 4 and a learning rate of $1 \times 10^{-4}$ for up to 100 epochs. For the $\kappa_\mathrm{L}$ task, we used a batch size of 1 along with a learning rate of $1 \times 10^{-3}$ and trained for up to 100 epochs. The detailed hyperparameters of KappaFormer are listed in {\color{blue}{Table S4}}. The overall forward propagation and dual-stage cross-domain training procedure are summarized in {\color{blue}{Algorithm 1}}.

\begin{myalgorithm}[t]{Forward propagation and dual-stage cross-domain training of KappaFormer}
\label{kappaformer_full}

\Require Crystal structure $\mathcal{M} = (\mathbf{A}, \mathbf{P}, \mathbf{L})$,
pre-training elastic dataset $\mathcal{D}_{\mathrm{pre}}$, fine-tuning $\kappa_\mathrm{L}$ dataset $\mathcal{D}_{\mathrm{ft}}$
\Ensure Trained model parameters $\boldsymbol{\theta}$, predicted properties $\hat{B}$, $\hat{G}$, $\hat{\kappa}_\mathrm{L}$

\State \textbf{Initialize} model parameters $\boldsymbol{\theta}$, vertex features $\mathbf{v}_i$ (Eq.~(\ref{vertex}))
\State \textbf{Compute} edge features $\mathbf{e}_{ij}$ (Eqs.~(\ref{edge_1})--(\ref{edge_2}))

\Function{Forward}{$\mathcal{M}=(\mathbf{A},\mathbf{P},\mathbf{L}); \boldsymbol{\theta}$}
    \State $\mathbf{x}_i^{(G)} \leftarrow \text{Compute initial embedding of each atom (Eq.~(\ref{initial}))}$
    \For{$t=1$ to $N$}
        \State $\mathbf{v}_{ij} \leftarrow \text{Compute value representation (Eq.~(\ref{value}))}$
        \State $\alpha_{ij} \leftarrow \text{Compute attention weight (Eq.~(\ref{weight}))}$
        \State $\mathbf{x}_i^{(A)} \leftarrow \text{Compute embedding after attention (Eq.~(\ref{x_A}))}$
    \EndFor

    \State $\mathbf{h}_i^{(A)}, \mathbf{h}_i^{(G)} \leftarrow \text{Compute invariant features (Eq.~(\ref{invariant}))}$

    \State $\mathbf{h}_i^{(S, \mathrm{Harm})} \leftarrow \mathbf{W}_{SH} \mathrm{GRU}^{(\mathrm{Harm})}(\mathbf{h}_i^{(G)}, \mathbf{h}_i^{(A)})$ 
    \State $\mathbf{h}_i^{(S, \mathrm{Anharm})} \leftarrow \mathbf{W}_{SA} \mathrm{GRU}^{(\mathrm{Anharm})}(\mathbf{h}_i^{(G)}, \mathbf{h}_i^{(S, \mathrm{Harm})})$ 

    \State $\mathbf{h}_i^{(M, \mathrm{Harm})} \leftarrow \text{Obtain harmonic features (Eq.~(\ref{expert}))}$ 
    \State $\mathbf{h}_i^{(M, \mathrm{Anharm})} \leftarrow \text{Obtain anharmonic features (Eq.~(\ref{expert}))}$ 

    \State $\mathbf{h}_{\mathcal{G}}^{(\xi)} \leftarrow \mathrm{Readout} (\{ \mathbf{h}_i^{(M, \xi)} \, \vert \, \mathbf{v}_i \in \mathcal{G}\})$

    \State $\hat{B} \leftarrow \mathrm{Dense}^{(B)} (\mathbf{h}_{\mathcal{G}}^{(\mathrm{Harm})})$
    \State $\hat{G} \leftarrow \mathrm{Dense}^{(G)} (\mathbf{h}_{\mathcal{G}}^{(\mathrm{Harm})})$
    \State $\hat{\gamma} \leftarrow \mathrm{Dense}^{(\gamma)}(\mathbf{h}_{\mathcal{G}}^{(\mathrm{Anharm})})$
    \State $\hat{\kappa}_L \leftarrow \log_{10} \hat{\kappa}_{\mathrm{L}}(\hat{B},\hat{G},\hat{\gamma} \mid  \boldsymbol{\theta})
  = { f(\hat{B},\hat{G} \mid \boldsymbol{\theta}) } + { g(\hat{\gamma} \mid \boldsymbol{\theta}) } + C$

    \State \Return $\hat{B}, \hat{G}, \hat{\kappa}_L$
\EndFunction

\Statex

\State \textbf{Stage I: Pre-training} \Comment{Harmonic branch} 

\For{epoch $=1...E_{\mathrm{pre}}$}
    \For{$(\mathcal{M}, B, G)$ in $\mathcal{D}_{\mathrm{pre}}$}
        \State $\hat{B}, \hat{G} \leftarrow \Call{Forward}{\mathcal{M}; \boldsymbol{\theta}}$
        \State $\mathcal{L}_{\mathrm{pre}} \leftarrow \mathcal{L}_H (\hat{B}, B) + \mathcal{L}_H (\hat{G}, G)$
        \State Update $\boldsymbol{\theta}$ by minimizing $\mathcal{L}_{\mathrm{pre}}$
    \EndFor
\EndFor

\State Save $\boldsymbol{\theta}^\star$

\Statex

\State \textbf{Stage II: Fine-tuning} \Comment{Anharmonic branch} 

\State Load $\boldsymbol{\theta}^\star$
\State Freeze backbone and harmonic layers

\For{epoch $=1...E_{\mathrm{ft}}$}
    \For{$(\mathcal{M}, B, G, \kappa_L)$ in $\mathcal{D}_{\mathrm{ft}}$}
        \State $\hat{B}, \hat{G}, \hat{\kappa}_L \leftarrow \Call{Forward}{\mathcal{M}; \boldsymbol{\theta}^\star}$
        \State $\mathcal{L}_{\mathrm{ft}} \leftarrow \mathcal{L}_H (\hat{B}, B) + \mathcal{L}_H (\hat{G}, G) + \alpha \mathcal{L}_{L1} (\hat{\kappa}_\mathrm{L}, \kappa_\mathrm{L})$
        \State Update trainable parameters only
    \EndFor
\EndFor

\State \Return $\boldsymbol{\theta}$

\end{myalgorithm}

\subsection{Theoretical framework of $\kappa_\mathrm{L}$}
Accurate $\kappa_\mathrm{L}$ values for CsNb$_2$Br$_9$, Cs$_2$AgI$_3$, and Cs$_6$CdSe$_4$ were calculated by iteratively solving the phonon BTE\cite{ShengBTE_2014, fugallo2013ab}
\begin{equation}
    \kappa_\mathrm{L}^{\alpha\beta}=\frac{1}{k_\mathrm{B}T^2\Omega N}\sum_\lambda f_0\left(f_0+1\right)\left(\hbar\omega_\lambda\right)^2\upsilon_\lambda^\alpha F_\lambda^\beta,
\end{equation}
where $\alpha$ and $\beta$ are the Cartesian indexes. $k_\mathrm{B}$, $T$, $\Omega$, $N$ are the Boltzmann constant, temperature, volume of the unit cell, and regular grid of \textit{q} points, respectively. $\lambda$ is a phonon mode including the branch index \textit{p} and wave vector \textit{q}, and $f_0$ is the phonon distribution function based on Bose-Einstein statistics. $\hbar$, $\omega_\lambda$, $\upsilon_\lambda^\alpha$ are the reduced Planck constant, phonon frequency, and phonon group velocity along the $\alpha$ direction, respectively. When only considering two- and three-phonon processes that contribute to scattering, the linearized BTE $F_\lambda$ takes the form $F_\lambda=\tau_\lambda^0(\upsilon_\lambda+\Delta_\lambda)$, where $\tau_\lambda^0$ and $\Delta_\lambda$ are the phonon lifetime of mode $\lambda$ and corrective term from iteration, respectively. 

\subsection{First-principles calculations}
All DFT-based first-principles calculations were performed using the projector-augmented wave (PAW) method,\cite{kresse1999ultrasoft} as implemented in the Vienna \textit{Ab initio} Simulation Package (VASP),\cite{kresse1996efficient} with data processing carried out via VASPKIT.\cite{wang2021vaspkit} The electronic exchange–correlation energy was treated using the Perdew–Burke–Ernzerhof (PBE) functional within the generalized gradient approximation (GGA).\cite{perdew1996generalized} Structural optimizations employed a plane-wave kinetic energy cutoff of 700 eV and a $\Gamma$-centered \textit{k}-point grid of 2$\pi$ $\times$ 0.02 $\mathrm{Å^{-1}}$ to sample the Brillouin zone, and were considered converged when the total energy change reached below $10^{-6}$ eV and all atomic forces were smaller than 0.001 eV/Å. The second-order IFCs were calculated by the finite displacement method\cite{baroni2001phonons} for calculations as implemented in the Phonopy package,\cite{phonopy-phono3py-JPCM, phonopy-phono3py-JPSJ} facilitating the evaluation of dynamic stability. In addition, the script \texttt{thirdorder.py}\cite{PhysRevB.86.174307} was employed to construct $1 \times 2 \times 1$, $1 \times 2 \times 1$, and $1 \times 1 \times 2$ supercells for CsNb$_2$Br$_9$, Cs$_2$AgI$_3$, and Cs$_6$CdSe$_4$, considering interactions up to the 15th nearest neighbors, which resulted in 4624, 1916, and 3232 supercells with displaced atoms for computing the third-order IFCs, respectively. Based on these computed second- and third-order IFCs, the convergent $\kappa_\mathrm{L}$ values of the selected materials were calculated with the ShengBTE package\cite{ShengBTE_2014} on $5 \times 7 \times 5$, $9 \times 13 \times 9$, and $7 \times 7 \times 7$ \textit{q}-point grids in reciprocal space. The crystal structures and ELF diagrams were visualized using VESTA.\cite{momma2011vesta} Moreover, the COHP was evaluated using the LOBSTER code\cite{dronskowski1993crystal} to characterize bonding, anti-bonding, and non-bonding interactions.

\section*{Data availability}
The data that support the findings of this study are available from the corresponding author upon reasonable request.

\acknowledgements
Mengfan Wu thanks Zekun Chen from Lawrence Berkeley National Laboratory and Rongkun Chen from Yunnan University for their fruitful discussions. This work is supported by the National Natural Science Foundation of China (No.~11935010), the National Key R\&D Program of China (No.~2023YFA1406900 and No.~2022YFA1404400), the Natural Science Foundation of Shanghai (No.~23ZR1481200), the Program of Shanghai Academic Research Leader (No.~23XD1423800), and the Opening Project of Shanghai Key Laboratory of Special Artificial Microstructure Materials and Technology.

\bibliography{mybib}

@article{butler2018machine,
  title={Machine learning for molecular and materials science},
  author={Butler, Keith T and Davies, Daniel W and Cartwright, Hugh and Isayev, Olexandr and Walsh, Aron},
  journal={Nature},
  volume={559},
  number={7715},
  pages={547--555},
  year={2018},
  publisher={Nature Publishing Group UK London}
}

@article{lookman2026materials,
  title={Materials informatics: Emergence to autonomous discovery in the age of {AI}},
  author={Lookman, Turab and Liu, YuJie and Gao, Zhibin},
  journal={Adv. Mater.},
  pages={e15941},
  year={2026},
  publisher={Wiley Online Library}
}

@article{ahlawat2026family,
  title={A family of large language models for materials research with insights into model adaptability in continued pretraining},
  author={Ahlawat, Dhruv and Mishra, Vaibhav and Singh, Somaditya and Zaki, Mohd and Bihani, Vaibhav and Grover, Hargun Singh and Mishra, Biswajit and Miret, Santiago and Mausam and Krishnan, NM Anoop},
  journal={Nat. Mach. Intell.},
  pages={1--14},
  year={2026},
  publisher={Nature Publishing Group UK London}
}

@article{li2023exploiting,
  title={Exploiting redundancy in large materials datasets for efficient machine learning with less data},
  author={Li, Kangming and Persaud, Daniel and Choudhary, Kamal and DeCost, Brian and Greenwood, Michael and Hattrick-Simpers, Jason},
  journal={Nat. Commun.},
  volume={14},
  number={1},
  pages={7283},
  year={2023},
  publisher={Nature Publishing Group UK London}
}

@article{cui2023atomic,
  title={Atomic positional embedding-based transformer model for predicting the density of states of crystalline materials},
  author={Cui, Yaning and Chen, Kang and Zhang, Lingyao and Wang, Haotian and Bai, Lei and Elliston, David and Ren, Wei},
  journal={J. Phys. Chem. Lett.},
  volume={14},
  number={35},
  pages={7924--7930},
  year={2023},
  publisher={ACS Publications}
}

@article{li2024high,
  title={High-throughput screening and machine learning classification of van der Waals dielectrics for 2{D} nanoelectronics},
  author={Li, Yuhui and Wan, Guolin and Zhu, Yongqian and Yang, Jingyu and Zhang, Yan-Fang and Pan, Jinbo and Du, Shixuan},
  journal={Nat. Commun.},
  volume={15},
  number={1},
  pages={9527},
  year={2024},
  publisher={Nature Publishing Group UK London}
}

@article{acharyya2022glassy,
  title={Glassy thermal conductivity in {Cs$_3$Bi$_2$I$_6$Cl$_3$} single crystal},
  author={Acharyya, Paribesh and Ghosh, Tanmoy and Pal, Koushik and Rana, Kewal Singh and Dutta, Moinak and Swain, Diptikanta and Etter, Martin and Soni, Ajay and Waghmare, Umesh V and Biswas, Kanishka},
  journal={Nat. Commun.},
  volume={13},
  number={1},
  pages={5053},
  year={2022},
  publisher={Nature Publishing Group UK London}
}

@article{wang2024interpretable,
  title={An interpretable formula for lattice thermal conductivity of crystals},
  author={Wang, Xiaoying and Shu, Guoyu and Zhu, Guimei and Wang, Jian-Sheng and Sun, Jun and Ding, Xiangdong and Li, Baowen and Gao, Zhibin},
  journal={Mater. Today Phys.},
  volume={48},
  pages={101549},
  year={2024},
  publisher={Elsevier}
}

@article{wu2023machine,
  title={Machine learning accelerated design of 2{D} covalent organic frame materials for thermoelectrics},
  author={Wu, Cheng-Wei and Li, Fan and Zeng, Yu-Jia and Zhao, Hongwei and Xie, Guofeng and Zhou, Wu-Xing and Liu, Qingquan and Zhang, Gang},
  journal={Appl. Surf. Sci.},
  volume={638},
  pages={157947},
  year={2023},
  publisher={Elsevier}
}

@article{wu2025ai,
  title={{AI}-empowered digital design of zeolites: Progress, challenges, and perspectives},
  author={Wu, Mengfan and Zhang, Shiyi and Ren, Jie},
  journal={APL Mater.},
  volume={13},
  number={2},
  pages={020601},
  year={2025},
  publisher={AIP Publishing}
}

@article{hu2025machine,
  title={Machine learning for the generative discovery of {{K}$^{+}$}-selective porous structures with aluminum sites},
  author={Hu, Jinbin and Wu, Mengfan and Ren, Jie},
  journal={Phys. Rev. Mater.},
  volume={9},
  number={7},
  pages={076002},
  year={2025},
  publisher={APS}
}

@incollection{lu2024crystal,
  title={Crystal Structure Prediction for Battery Materials},
  author={Lu, Ziheng and Zhu, Bonan},
  booktitle={Computational Design of Battery Materials},
  pages={187--210},
  year={2024},
  publisher={Springer}
}

@article{jain2013commentary,
  title={Commentary: The {Materials Project}: A materials genome approach to accelerating materials innovation},
  author={Jain, Anubhav and Ong, Shyue Ping and Hautier, Geoffroy and Chen, Wei and Richards, William Davidson and Dacek, Stephen and Cholia, Shreyas and Gunter, Dan and Skinner, David and Ceder, Gerbrand and others},
  journal={APL Mater.},
  volume={1},
  number={1},
  pages={011002},
  year={2013},
  publisher={AIP Publishing}
}

@article{saal2013materials,
  title={Materials design and discovery with high-throughput density functional theory: The {Open Quantum Materials Database (OQMD)}},
  author={Saal, James E and Kirklin, Scott and Aykol, Muratahan and Meredig, Bryce and Wolverton, Christopher},
  journal={Jom},
  volume={65},
  pages={1501--1509},
  year={2013},
  publisher={Springer}
}

@article{kirklin2015open,
  title={The {Open Quantum Materials Database (OQMD)}: Assessing the accuracy of {DFT} formation energies},
  author={Kirklin, Scott and Saal, James E and Meredig, Bryce and Thompson, Alex and Doak, Jeff W and Aykol, Muratahan and R{\"u}hl, Stephan and Wolverton, Chris},
  journal={Npj Comput. Mater.},
  volume={1},
  number={1},
  pages={1--15},
  year={2015},
  publisher={Nature Publishing Group}
}

@article{li2025high,
  title={High-throughput computational framework for high-order anharmonic thermal transport in cubic and tetragonal crystals},
  author={Li, Zhi and Lee, Huiju and Wolverton, Chris and Xia, Yi},
  journal={Npj Comput. Mater.},
  volume={12},
  number={1},
  pages={51},
  year={2026},
  publisher={Nature Publishing Group UK London}
}

@article{zheng2024unravelling,
  title={Unravelling ultralow thermal conductivity in perovskite {Cs$_2$AgBiBr$_6$}: Dominant wave-like phonon tunnelling and strong anharmonicity},
  author={Zheng, Jiongzhi and Lin, Changpeng and Lin, Chongjia and Hautier, Geoffroy and Guo, Ruiqiang and Huang, Baoling},
  journal={Npj Comput. Mater.},
  volume={10},
  number={1},
  pages={30},
  year={2024},
  publisher={Nature Publishing Group UK London}
}

@article{madani2025accelerating,
  title={Accelerating materials property prediction via a hybrid Transformer Graph framework that leverages four body interactions},
  author={Madani, Mohammad and Lacivita, Valentina and Shin, Yongwoo and Tarakanova, Anna},
  journal={Npj Comput. Mater.},
  volume={11},
  number={1},
  pages={15},
  year={2025},
  publisher={Nature Publishing Group UK London}
}

@article{guo2025generative,
  title={Generative deep learning for predicting ultrahigh lattice thermal conductivity materials},
  author={Guo, Liben and Liu, Yuanbin and Chen, Zekun and Yang, Hongao and Donadio, Davide and Cao, Bingyang},
  journal={Npj Comput. Mater.},
  volume={11},
  number={1},
  pages={97},
  year={2025},
  publisher={Nature Publishing Group UK London}
}

@article{dong2025accurate,
  title={Accurate piezoelectric tensor prediction with equivariant attention tensor graph neural network},
  author={Dong, L and Zhang, X and Yang, Z and Shen, L and Lu, Y},
  journal={Npj Comput. Mater.},
  volume={11},
  number={1},
  pages={63},
  year={2025},
  publisher={Nature Publishing Group UK London}
}

@article{wu2025hierarchy,
  title={Hierarchy-boosted funnel learning for identifying semiconductors with ultralow lattice thermal conductivity},
  author={Wu, Mengfan and Yan, Shenshen and Ren, Jie},
  journal={Npj Comput. Mater.},
  volume={11},
  number={1},
  pages={106},
  year={2025},
  publisher={Nature Publishing Group UK London}
}

@article{choudhary2021atomistic,
  title={Atomistic line graph neural network for improved materials property predictions},
  author={Choudhary, Kamal and DeCost, Brian},
  journal={Npj Comput. Mater.},
  volume={7},
  number={1},
  pages={185},
  year={2021},
  publisher={Nature Publishing Group UK London}
}

@article{li2025probing,
  title={Probing the limit of heat transfer in inorganic crystals with deep learning},
  author={Li, Jielan and Chen, Zekun and Wang, Qian and Yang, Han and Lu, Ziheng and Li, Guanzhi and Chen, Shuizhou and Zhu, Yu and Liu, Xixian and Tan, Junfu and others},
  journal={arXiv preprint arXiv:2503.11568},
  year={2025}
}

@article{curtarolo2012aflow,
  title={{AFLOW}: An automatic framework for high-throughput materials discovery},
  author={Curtarolo, Stefano and Setyawan, Wahyu and Hart, Gus LW and Jahnatek, Michal and Chepulskii, Roman V and Taylor, Richard H and Wang, Shidong and Xue, Junkai and Yang, Kesong and Levy, Ohad and others},
  journal={Comput. Mater. Sci.},
  volume={58},
  pages={218--226},
  year={2012},
  publisher={Elsevier}
}

@article{choudhary2020joint,
  title={The joint automated repository for various integrated simulations ({JARVIS}) for data-driven materials design},
  author={Choudhary, Kamal and Garrity, Kevin F and Reid, Andrew CE and DeCost, Brian and Biacchi, Adam J and Hight Walker, Angela R and Trautt, Zachary and Hattrick-Simpers, Jason and Kusne, A Gilad and Centrone, Andrea and others},
  journal={Npj Comput. Mater.},
  volume={6},
  number={1},
  pages={173},
  year={2020},
  publisher={Nature Publishing Group UK London}
}

@article{cheng2026artificial,
  title={Artificial intelligence-driven approaches for materials design and discovery},
  author={Cheng, Mouyang and Fu, Chu-Liang and Okabe, Ryotaro and Chotrattanapituk, Abhijatmedhi and Boonkird, Artittaya and Hung, Nguyen Tuan and Li, Mingda},
  journal={Nat. Mater.},
  pages={1--17},
  year={2026},
  publisher={Nature Publishing Group UK London}
}

@article{qian2021phonon,
  title={Phonon-engineered extreme thermal conductivity materials},
  author={Qian, Xin and Zhou, Jiawei and Chen, Gang},
  journal={Nat. Mater.},
  volume={20},
  number={9},
  pages={1188--1202},
  year={2021},
  publisher={Nature Publishing Group UK London}
}

@article{li2022deep,
  title={Deep-learning density functional theory Hamiltonian for efficient ab initio electronic-structure calculation},
  author={Li, He and Wang, Zun and Zou, Nianlong and Ye, Meng and Xu, Runzhang and Gong, Xiaoxun and Duan, Wenhui and Xu, Yong},
  journal={Nat. Comput. Sci.},
  volume={2},
  number={6},
  pages={367--377},
  year={2022},
  publisher={Nature Publishing Group US New York}
}

@article{riebesell2025framework,
  title={A framework to evaluate machine learning crystal stability predictions},
  author={Riebesell, Janosh and Goodall, Rhys EA and Benner, Philipp and Chiang, Yuan and Deng, Bowen and Ceder, Gerbrand and Asta, Mark and Lee, Alpha A and Jain, Anubhav and Persson, Kristin A},
  journal={Nat. Mach. Intell.},
  volume={7},
  number={6},
  pages={836--847},
  year={2025},
  publisher={Nature Publishing Group UK London}
}

@article{wen2024equivariant,
  title={An equivariant graph neural network for the elasticity tensors of all seven crystal systems},
  author={Wen, Mingjian and Horton, Matthew K and Munro, Jason M and Huck, Patrick and Persson, Kristin A},
  journal={Digit. Discov.},
  volume={3},
  number={5},
  pages={869--882},
  year={2024},
  publisher={Royal Society of Chemistry}
}

@article{klimova2025symmetry,
  title={Symmetry-aware equivariant network for discovering {SHG}-active materials},
  author={Klimova, Liudmila A and Trofimov, Ivan S and Jin, Wenqi and Song, Qigang and Arsenin, Aleksey and Xie, Congwei and Kruglov, Ivan A and Volkov, Valentyn},
  journal={Adv. Funct. Mater.},
  pages={e23683},
  year={2025},
  publisher={Wiley Online Library}
}

@article{li2024phonon,
  title={Phonon Coherence in Bismuth-Halide Perovskite {Cs$_3$Bi$_2$Br$_9$} With Ultralow Thermal Conductivity},
  author={Li, Yongheng and Li, Xiang and Wei, Bin and Liu, Juanjuan and Pan, Feihao and Wang, Hongliang and Cheng, Peng and Zhang, Hongxia and Xu, Daye and Bao, Wei and others},
  journal={Adv. Funct. Mater.},
  volume={34},
  number={52},
  pages={2411152},
  year={2024},
  publisher={Wiley Online Library}
}

@article{wu2020comprehensive,
  title={A comprehensive survey on graph neural networks},
  author={Wu, Zonghan and Pan, Shirui and Chen, Fengwen and Long, Guodong and Zhang, Chengqi and Yu, Philip S},
  journal={IEEE Trans. Neural Netw. Learn. Syst.},
  volume={32},
  number={1},
  pages={4--24},
  year={2020},
  publisher={IEEE}
}

@inproceedings{satorras2021n,
  title={{E}(n) equivariant graph neural networks},
  author={Satorras, V{\i}ctor Garcia and Hoogeboom, Emiel and Welling, Max},
  booktitle={International Conference on Machine Learning},
  pages={9323--9332},
  year={2021},
  organization={PMLR}
}

@article{vaswani2017attention,
  title={Attention is all you need},
  author={Vaswani, Ashish and Shazeer, Noam and Parmar, Niki and Uszkoreit, Jakob and Jones, Llion and Gomez, Aidan N and Kaiser, {\L}ukasz and Polosukhin, Illia},
  journal={Advances in Neural Information Processing Systems},
  volume={30},
  year={2017}
}

@article{chen2025softening,
  title={Softening of Vibrational Modes and Anharmonicity Induced Thermal Conductivity Reduction in a-{S}i: {H} at High Temperatures},
  author={Chen, Zhuo and Yuan, Yuejin and Wang, Yanzhou and Ying, Penghua and Li, Shouhang and Shao, Cheng and Ding, Wenyang and Zhang, Gang and An, Meng},
  journal={Adv. Electron. Mater.},
  volume={11},
  number={13},
  pages={2500104},
  year={2025},
  publisher={Wiley Online Library}
}

@article{song2026role,
  title={Role of strong anisotropy and high-order anharmonicity in the phonon thermal transport of pentagonal {PdX$_2$ (X= S, Se, Te)} monolayers},
  author={Song, Xiefei and Tang, Gang and Li, Wenzhong and Qin, Mengran and Wang, Ning and He, Yao},
  journal={Phys. Rev. B},
  volume={113},
  number={3},
  pages={035430},
  year={2026},
  publisher={APS}
}

@article{li2021optical,
  title={Optical phonon dominated heat transport: A first-principles thermal conductivity study of {BaSnS$_2$}},
  author={Li, Zhi and Xie, Hongyao and Hao, Shiqiang and Xia, Yi and Su, Xianli and Kanatzidis, Mercouri G and Wolverton, Christopher and Tang, Xinfeng},
  journal={Phys. Rev. B},
  volume={104},
  number={24},
  pages={245209},
  year={2021},
  publisher={APS}
}

@article{zheng2022anharmonicity,
  title={Anharmonicity-induced phonon hardening and phonon transport enhancement in crystalline perovskite {BaZrO$_3$}},
  author={Zheng, Jiongzhi and Shi, Dongliang and Yang, Yuewang and Lin, Chongjia and Huang, He and Guo, Ruiqiang and Huang, Baoling},
  journal={Phys. Rev. B},
  volume={105},
  number={22},
  pages={224303},
  year={2022},
  publisher={APS}
}

@article{beltukov2013ioffe,
  title={Ioffe-Regel criterion and diffusion of vibrations in random lattices},
  author={Beltukov, YM and Kozub, VI and Parshin, DA},
  journal={Phys. Rev. B},
  volume={87},
  number={13},
  pages={134203},
  year={2013},
  publisher={APS}
}

@article{wang2024revisiting,
  title={Revisiting lattice thermal conductivity of {CsCl}: The crucial role of quartic anharmonicity},
  author={Wang, Xiaoying and Feng, Minxuan and Xia, Yi and Sun, Jun and Ding, Xiangdong and Li, Baowen and Gao, Zhibin},
  journal={Appl. Phys. Lett.},
  volume={124},
  number={17},
  pages={172201},
  year={2024},
  publisher={AIP Publishing}
}

@article{padture2002thermal,
  title={Thermal barrier coatings for gas-turbine engine applications},
  author={Padture, Nitin P and Gell, Maurice and Jordan, Eric H},
  journal={Science},
  volume={296},
  number={5566},
  pages={280--284},
  year={2002},
  publisher={American Association for the Advancement of Science}
}

@article{song2018two,
  title={Two-dimensional materials for thermal management applications},
  author={Song, Houfu and Liu, Jiaman and Liu, Bilu and Wu, Junqiao and Cheng, Hui-Ming and Kang, Feiyu},
  journal={Joule},
  volume={2},
  number={3},
  pages={442--463},
  year={2018},
  publisher={Elsevier}
}

@article{xiang2022thermal,
  title={Thermal transport in lithium-ion battery: A micro perspective for thermal management},
  author={Xiang, Changqing and Wu, Cheng-Wei and Zhou, Wu-Xing and Xie, Guofeng and Zhang, Gang},
  journal={Front. Phys.},
  volume={17},
  pages={1--11},
  year={2022},
  publisher={Springer}
}

@article{nolas1999skutterudites,
  title={Skutterudites: A phonon-glass-electron crystal approach to advanced thermoelectric energy conversion applications},
  author={Nolas, GS and Morelli, DT and Tritt, Terry M},
  journal={Annu. Rev. Mater. Sci.},
  volume={29},
  number={1},
  pages={89--116},
  year={1999},
  publisher={Annual Reviews 4139 El Camino Way, PO Box 10139, Palo Alto, CA 94303-0139, USA}
}

@article{snyder2004disordered,
  title={Disordered zinc in {Zn\textsubscript{4}Sb\textsubscript{3}} with phonon-glass and electron-crystal thermoelectric properties},
  author={Snyder, G Jeffrey and Christensen, Mogens and Nishibori, Eiji and Caillat, Thierry and Iversen, Bo Brummerstedt},
  journal={Nat. Mater.},
  volume={3},
  number={7},
  pages={458--463},
  year={2004},
  publisher={Nature Publishing Group UK London}
}

@article{balandin2011thermal,
  title={Thermal properties of graphene and nanostructured carbon materials},
  author={Balandin, Alexander A},
  journal={Nat. Mater.},
  volume={10},
  number={8},
  pages={569--581},
  year={2011},
  publisher={Nature Publishing Group UK London}
}

@article{snyder2008complex,
  title={Complex thermoelectric materials},
  author={Snyder, G Jeffrey and Toberer, Eric S},
  journal={Nat. Mater.},
  volume={7},
  number={2},
  pages={105--114},
  year={2008},
  publisher={Nature Publishing Group UK London}
}

@article{wei2016intrinsic,
  title={The intrinsic thermal conductivity of SnSe},
  author={Wei, Pai-Chun and Bhattacharya, S and He, J and Neeleshwar, S and Podila, R and Chen, YY and Rao, AM},
  journal={Nature},
  volume={539},
  number={7627},
  pages={E1--E2},
  year={2016},
  publisher={Nature Publishing Group UK London}
}

@article{ShengBTE_2014,
   author={Wu Li and Jes\'us Carrete and Nebil A. Katcho and Natalio Mingo},
   title={{ShengBTE}: A solver of the {B}oltzmann transport equation for phonons},
   journal={Comput. Phys. Commun.},
   volume={185},
   pages={1747–1758},
   year={2014}
}

@article{merchant2023scaling,
  title={Scaling deep learning for materials discovery},
  author={Merchant, Amil and Batzner, Simon and Schoenholz, Samuel S and Aykol, Muratahan and Cheon, Gowoon and Cubuk, Ekin Dogus},
  journal={Nature},
  volume={624},
  number={7990},
  pages={80--85},
  year={2023},
  publisher={Nature Publishing Group UK London}
}

@article{jaafreh2021lattice,
  title={Lattice thermal conductivity: An accelerated discovery guided by machine learning},
  author={Jaafreh, Russlan and Kang, Yoo Seong and Hamad, Kotiba},
  journal={ACS Appl. Mater. Interfaces},
  volume={13},
  number={48},
  pages={57204--57213},
  year={2021},
  publisher={ACS Publications}
}

@article{luo2023predicting,
  title={Predicting lattice thermal conductivity via machine learning: A mini review},
  author={Luo, Yufeng and Li, Mengke and Yuan, Hongmei and Liu, Huijun and Fang, Ying},
  journal={Npj Comput. Mater.},
  volume={9},
  number={1},
  pages={4},
  year={2023},
  publisher={Nature Publishing Group UK London}
}

@article{chen2019machine,
  title={Machine learning models for the lattice thermal conductivity prediction of inorganic materials},
  author={Chen, Lihua and Tran, Huan and Batra, Rohit and Kim, Chiho and Ramprasad, Rampi},
  journal={Comput. Mater. Sci.},
  volume={170},
  pages={109155},
  year={2019},
  publisher={Elsevier}
}

@inproceedings{ma2018modeling,
  title={Modeling task relationships in multi-task learning with {Multi-gate Mixture-of-Experts}},
  author={Ma, Jiaqi and Zhao, Zhe and Yi, Xinyang and Chen, Jilin and Hong, Lichan and Chi, Ed H},
  booktitle={Proceedings of the 24th ACM SIGKDD International Conference on Knowledge Discovery and Data Mining},
  pages={1930--1939},
  year={2018}
}

@article{schutt2017schnet,
  title={Schnet: A continuous-filter convolutional neural network for modeling quantum interactions},
  author={Sch{\"u}tt, Kristof and Kindermans, Pieter-Jan and Sauceda Felix, Huziel Enoc and Chmiela, Stefan and Tkatchenko, Alexandre and M{\"u}ller, Klaus-Robert},
  journal={Advances in Neural Information Processing Systems},
  volume={30},
  year={2017}
}

@article{paszke2019pytorch,
  title={Pytorch: An imperative style, high-performance deep learning library},
  author={Paszke, Adam and Gross, Sam and Massa, Francisco and Lerer, Adam and Bradbury, James and Chanan, Gregory and Killeen, Trevor and Lin, Zeming and Gimelshein, Natalia and Antiga, Luca and others},
  journal={Advances in Neural Information Processing Systems},
  volume={32},
  year={2019}
}

@article{xie2018crystal,
  title={Crystal graph convolutional neural networks for an accurate and interpretable prediction of material properties},
  author={Xie, Tian and Grossman, Jeffrey C},
  journal={Phys. Rev. Lett.},
  volume={120},
  number={14},
  pages={145301},
  year={2018},
  publisher={APS}
}

@article{pailhes2014localization,
  title={Localization of propagative phonons in a perfectly crystalline solid},
  author={Pailh{\`e}s, St{\'e}phane and Euchner, H and Giordano, Valentina M and Debord, R{\'e}gis and Assy, A and Gom{\`e}s, S and Bosak, A and Machon, Denis and Paschen, S and De Boissieu, M},
  journal={Phys. Rev. Lett.},
  volume={113},
  number={2},
  pages={025506},
  year={2014},
  publisher={APS}
}

@article{tadano2015impact,
  title={Impact of rattlers on thermal conductivity of a thermoelectric clathrate: A first-principles study},
  author={Tadano, Terumasa and Gohda, Yoshihiro and Tsuneyuki, Shinji},
  journal={Phys. Rev. Lett.},
  volume={114},
  number={9},
  pages={095501},
  year={2015},
  publisher={APS}
}

@article{morelli2008intrinsically,
  title={Intrinsically minimal thermal conductivity in cubic {I--V--VI$_2$} semiconductors},
  author={Morelli, DT and Jovovic, V and Heremans, JP},
  journal={Phys. Rev. Lett.},
  volume={101},
  number={3},
  pages={035901},
  year={2008},
  publisher={APS}
}

@inproceedings{
Gasteiger2020Directional,
title={Directional Message Passing for Molecular Graphs},
author={Gasteiger, Johannes and Gro{\ss}, Janek and G{\"u}nnemann, Stephan},
booktitle={International Conference on Learning Representations},
year={2020}
}

@inproceedings{
liao2023equiformer,
  title={Equiformer: Equivariant graph attention transformer for 3{D} atomistic graphs},
  author={Liao, Yi-Lun and Smidt, Tess},
  booktitle={International Conference on Learning Representations},
  year={2023}
}

@article{ying2021transformers,
  title={Do transformers really perform badly for graph representation?},
  author={Ying, Chengxuan and Cai, Tianle and Luo, Shengjie and Zheng, Shuxin and Ke, Guolin and He, Di and Shen, Yanming and Liu, Tie-Yan},
  journal={Advances in Neural Information Processing Systems},
  volume={34},
  pages={28877--28888},
  year={2021}
}

@inproceedings{cho2014learning,
  title={Learning phrase representations using RNN encoder--decoder for statistical machine translation},
  author={Cho, Kyunghyun and Van Merri{\"e}nboer, Bart and Gul{\c{c}}ehre, {\c{C}}a{\u{g}}lar and Bahdanau, Dzmitry and Bougares, Fethi and Schwenk, Holger and Bengio, Yoshua},
  booktitle = {Proceedings of the 2014 Conference on Empirical Methods in Natural Language Processing},
  pages={1724--1734},
  year={2014}
}

@article{slack1973nonmetallic,
  title={Nonmetallic crystals with high thermal conductivity},
  author={Slack, Glen A},
  journal={J. Phys. Chem. Solids},
  volume={34},
  number={2},
  pages={321--335},
  year={1973},
  publisher={Elsevier}
}

@article{anderson1963simplified,
  title={A simplified method for calculating the {D}ebye temperature from elastic constants},
  author={Anderson, Orson L},
  journal={J. Phys. Chem. Solids},
  volume={24},
  number={7},
  pages={909--917},
  year={1963},
  publisher={Elsevier}
}

@article{jia2017lattice,
  title={Lattice thermal conductivity evaluated using elastic properties},
  author={Jia, Tiantian and Chen, Gang and Zhang, Yongsheng},
  journal={Phys. Rev. B},
  volume={95},
  number={15},
  pages={155206},
  year={2017},
  publisher={APS}
}

@inproceedings{li2016gated,
  author       = {Yujia Li and
                  Daniel Tarlow and
                  Marc Brockschmidt and
                  Richard S. Zemel},
  title        = {Gated Graph Sequence Neural Networks},
  booktitle = {International Conference on Learning Representations},
  year         = {2016}
}

@article{huber1992robust,
  title={Robust estimation of a location parameter},
  author={Huber, Peter J},
  journal={Ann. Math. Stat.},
  volume={35},
  pages={73--101},
  year={1964}
}

@article{schutt2018schnet,
  title={Schnet--a deep learning architecture for molecules and materials},
  author={Sch{\"u}tt, Kristof T and Sauceda, Huziel E and Kindermans, P-J and Tkatchenko, Alexandre and M{\"u}ller, K-R},
  journal={J. Chem. Phys.},
  volume={148},
  number={24},
  pages={241722},
  year={2018},
  publisher={AIP Publishing}
}

@article{chen2019graph,
  title={Graph networks as a universal machine learning framework for molecules and crystals},
  author={Chen, Chi and Ye, Weike and Zuo, Yunxing and Zheng, Chen and Ong, Shyue Ping},
  journal={Chem. Mater.},
  volume={31},
  number={9},
  pages={3564--3572},
  year={2019},
  publisher={ACS Publications}
}

@article{omee2022scalable,
  title={Scalable deeper graph neural networks for high-performance materials property prediction},
  author={Omee, Sadman Sadeed and Louis, Steph-Yves and Fu, Nihang and Wei, Lai and Dey, Sourin and Dong, Rongzhi and Li, Qinyang and Hu, Jianjun},
  journal={Patterns},
  volume={3},
  number={5},
  year={2022},
  publisher={Elsevier}
}

@article{yan2022periodic,
  title={Periodic graph transformers for crystal material property prediction},
  author={Yan, Keqiang and Liu, Yi and Lin, Yuchao and Ji, Shuiwang},
  journal={Advances in Neural Information Processing Systems},
  volume={35},
  pages={15066--15080},
  year={2022}
}

@article{van2008visualizing,
  title={Visualizing data using t-{SNE}.},
  author={Van der Maaten, Laurens and Hinton, Geoffrey},
  journal={J. Mach. Learn. Res.},
  volume={9},
  number={11},
  pages={2579--2605},
  year={2008}
}

@article{li2025long,
  title={Long-Range Anion Correlations Mediating Dynamic Anharmonicity and Contributing to Glassy Thermal Conductivity in Well-Ordered {K$_2$Ag$_4$Se$_3$}},
  author={Li, Fan and Liu, Xin and Yang, Jiawei and Wang, Xin-Ye and Yang, Yi-Chang and Ma, Ni and Chen, Ling and Wu, Li-Ming},
  journal={Small},
  volume={21},
  number={4},
  pages={2409524},
  year={2025},
  publisher={Wiley Online Library}
}

@article{mukhopadhyay2018two,
  title={Two-channel model for ultralow thermal conductivity of crystalline {Tl$_3$VSe$_4$}},
  author={Mukhopadhyay, Saikat and Parker, David S and Sales, Brian C and Puretzky, Alexander A and McGuire, Michael A and Lindsay, Lucas},
  journal={Science},
  volume={360},
  number={6396},
  pages={1455--1458},
  year={2018},
  publisher={American Association for the Advancement of Science}
}

@article{dronskowski1993crystal,
  title={Crystal orbital Hamilton populations ({COHP}): Energy-resolved visualization of chemical bonding in solids based on density-functional calculations},
  author={Dronskowski, Richard and Bloechl, Peter E},
  journal={J. Phys. Chem.},
  volume={97},
  number={33},
  pages={8617--8624},
  year={1993},
  publisher={ACS Publications}
}

@article{kresse1999ultrasoft,
  title={From ultrasoft pseudopotentials to the projector augmented-wave method},
  author={Kresse, Georg and Joubert, Daniel},
  journal={Phys. Rev. B},
  volume={59},
  number={3},
  pages={1758--1775},
  year={1999},
  publisher={APS}
}

@article{kresse1996efficient,
  title={Efficient iterative schemes for ab initio total-energy calculations using a plane-wave basis set},
  author={Kresse, Georg and Furthm{\"u}ller, J{\"u}rgen},
  journal={Phys. Rev. B},
  volume={54},
  number={16},
  pages={11169},
  year={1996},
  publisher={APS}
}

@article{wang2021vaspkit,
  title={{VASPKIT}: A user-friendly interface facilitating high-throughput computing and analysis using {VASP} code},
  author={Wang, Vei and Xu, Nan and Liu, Jin-Cheng and Tang, Gang and Geng, Wen-Tong},
  journal={Comput. Phys. Commun.},
  volume={267},
  pages={108033},
  year={2021},
  publisher={Elsevier}
}

@article{perdew1996generalized,
  title={Generalized gradient approximation made simple},
  author={Perdew, John P and Burke, Kieron and Ernzerhof, Matthias},
  journal={Phys. Rev. Lett.},
  volume={77},
  number={18},
  pages={3865},
  year={1996},
  publisher={APS}
}

@article{baroni2001phonons,
  title={Phonons and related crystal properties from density-functional perturbation theory},
  author={Baroni, Stefano and De Gironcoli, Stefano and Dal Corso, Andrea and Giannozzi, Paolo},
  journal={Rev. Mod. Phys.},
  volume={73},
  number={2},
  pages={515},
  year={2001},
  publisher={APS}
}

@article{phonopy-phono3py-JPCM,
  author  = {Togo, Atsushi and Chaput, Laurent and Tadano, Terumasa and Tanaka, Isao},
  title   = {Implementation strategies in phonopy and phono3py},
  journal = {J. Phys. Condens. Matter},
  volume  = {35},
  number  = {35},
  pages   = {353001},
  year    = {2023}
}

@article{phonopy-phono3py-JPSJ,
  author  = {Togo, Atsushi},
  title   = {First-principles Phonon Calculations with Phonopy and Phono3py},
  journal = {J. Phys. Soc. Jpn.},
  volume  = {92},
  number  = {1},
  pages   = {012001},
  year    = {2023}
}

@article{PhysRevB.86.174307,
   author={Wu Li and Lucas Lindsay and David A. Broido and Derek A. Stewart and Natalio Mingo},
   title={Thermal conductivity of bulk and nanowire {Mg\textsubscript{2}Si\textsubscript{x}Sn\textsubscript{1-x}} alloys from first principles},
   journal={Phys. Rev. B},
   volume={86},
   pages={174307},
   year = {2012}
}

@article{momma2011vesta,
  title={{VESTA} 3 for three-dimensional visualization of crystal, volumetric and morphology data},
  author={Momma, Koichi and Izumi, Fujio},
  journal={J. Appl. Crystallogr.},
  volume={44},
  number={6},
  pages={1272--1276},
  year={2011},
  publisher={International Union of Crystallography}
}

@article{fugallo2013ab,
  title={Ab initio variational approach for evaluating lattice thermal conductivity},
  author={Fugallo, Giorgia and Lazzeri, Michele and Paulatto, Lorenzo and Mauri, Francesco},
  journal={Phys. Rev. B},
  volume={88},
  number={4},
  pages={045430},
  year={2013},
  publisher={APS}
}

@inproceedings{
    equiformer_v2,
    title={{EquiformerV2: Improved equivariant transformer for scaling to higher-degree representations}}, 
    author={Yi-Lun Liao and Brandon Wood and Abhishek Das and Tess Smidt},
    booktitle={International Conference on Learning Representations},
    year={2024}
}

@inproceedings{
    escn,
    title={{Reducing SO(3) convolutions to SO(2) for efficient equivariant GNNs}},
    author={Passaro, Saro and Zitnick, C Lawrence},
    booktitle={International Conference on Machine Learning},
    year={2023}
}

@article{geiger2022e3nn,
  title={e3nn: Euclidean neural networks},
  author={Geiger, Mario and Smidt, Tess},
  journal={arXiv preprint arXiv:2207.09453},
  year={2022}
}

@book{goodman2000representations,
  title={Representations and invariants of the classical groups},
  author={Goodman, Roe and Wallach, Nolan R},
  year={2000},
  publisher={Cambridge University Press}
}

@book{griffiths2018introduction,
  title={Introduction to quantum mechanics},
  author={Griffiths, David J and Schroeter, Darrell F},
  year={2018},
  publisher={Cambridge university press}
}

\end{document}